\documentclass[onecolumn,floatfix,superscriptaddress,showpacs,showkeys,nofootinbib,preprint]{revtex4}

\usepackage{epsfig}
\usepackage{amssymb,latexsym,amsmath}
\newcommand{\eq}[1]{\begin{align} #1 \end{align}}

\begin{document}

\title{Multiplicity fluctuations \\
       in relativistic nuclear collisions: \\
       statistical model versus experimental data
}

 \author{V.V. Begun}
 \affiliation{Museo Storico della Fisica e Centro Studi e Ricerche
 Enrico Fermi, Rome, Italy}
 \affiliation{Bogolyubov Institute for Theoretical Physics, Kiev, Ukraine}

 \author{M. Ga\'zdzicki}
 \affiliation{Institut f\"ur Kernphysik, University of Frankfurt, Frankfurt, Germany}
 \affiliation{\'Swi\c{e}tokrzyska Academy, Kielce, Poland}

 \author{M.I. Gorenstein}
 \affiliation{Bogolyubov Institute for Theoretical Physics, Kiev, Ukraine}
 \affiliation{Frankfurt Institute for Advanced Studies, Frankfurt,Germany}

 \author{M. Hauer}
 \affiliation{Helmholtz Research School, University of Frankfurt, Frankfurt, Germany}
 \affiliation{University of Cape Town, Cape Town, South Africa}

 \author{V.P. Konchakovski}
 \affiliation{Bogolyubov Institute for Theoretical Physics, Kiev, Ukraine}
 \affiliation{Helmholtz Research School, University of Frankfurt, Frankfurt, Germany}

 \author{B. Lungwitz}
 \affiliation{Institut f\"ur Kernphysik, University of Frankfurt, Frankfurt, Germany}

\begin{abstract}
The multiplicity distributions of hadrons produced
in central nucleus-nucleus collisions are studied
within the hadron-resonance gas model in the large volume limit.
The microscopic correlator method is used to enforce
conservation of three charges -- baryon number, electric charge, and
strangeness -- in the canonical  ensemble. In addition, in the
micro-canonical ensemble  energy conservation is included.
An analytical method is used to account for resonance decays.
The multiplicity distributions and the scaled variances
for negatively, positively, and all charged hadrons
are calculated along the chemical freeze-out line of central Pb+Pb
(Au+Au) collisions from SIS to LHC  energies.
Predictions obtained within
different statistical ensembles
are compared with the preliminary NA49 experimental results
on central Pb+Pb collisions in the SPS energy range.
The measured fluctuations are significantly narrower than
the Poisson ones and clearly favor
expectations for  the micro-canonical ensemble.
Thus this is a first  observation of the
recently predicted  suppression of the multiplicity fluctuations
in relativistic gases in the thermodynamical limit due
to conservation laws.
\end{abstract}



\maketitle

\section{Introduction}
For more than 50 years statistical models of strong interactions
\cite{fermi,landau,hagedorn} have served  as an important tool to
investigate high energy nuclear collisions. The main subject of
the past study
has been the mean multiplicity of produced
hadrons (see e.g. Refs. \cite{stat1,FOC,FOP,pbm}). Only recently, due
to a rapid development of
experimental
techniques,
first measurements of fluctuations of
particle multiplicity \cite{fluc-mult} and transverse momenta \cite{fluc-pT}
were performed.
The growing interest
in the study of fluctuations in strong interactions (see e.g., reviews
\cite{fluc1}) is motivated
by expectations of anomalies
in the vicinity of the onset of deconfinement \cite{ood} and
in the case when the expanding system goes through
the transition line between
the quark-gluon plasma and the hadron gas \cite{fluc2}.
In particular, a critical point
of strongly interacting matter may be signaled by a characteristic
power-law pattern in fluctuations \cite{fluc3}. Apart from being an
important tool in an effort to study the critical behavior,
the study of fluctuations in the statistical hadronization model
constitutes a further test of its validity.
In this paper we make, for the first time, predictions for the
multiplicity fluctuations in central collisions of heavy nuclei calculated within the
micro-canonical formulation of the hadron-resonance gas model.
Fluctuations are quantified by the ratio of the variance of the
multiplicity distribution and its mean value, the so-called scaled
variance. The model calculations are compared with the
corresponding preliminary results \cite{NA49} of NA49 on central
Pb+Pb collisions at the CERN SPS energies.

There is a qualitative difference in the properties of the mean
multiplicity and the scaled variance of multiplicity distribution
in statistical models. In the case of the mean multiplicity
results obtained with the grand canonical ensemble (GCE), canonical
ensemble (CE), and micro-canonical ensemble (MCE)  approach  each
other in the large volume limit. One refers here to the
thermodynamical equivalence of the statistical ensembles. It was
recently found  \cite{CE,res} that corresponding results for the
scaled variance are different in different ensembles, and thus the
scaled variance is sensitive to conservation laws obeyed by a
statistical system. The differences are preserved  in the
thermodynamic limit.

The paper is organized as follows. In Section II the microscopic
correlators for a relativistic quantum gas are calculated in the MCE
in the thermodynamical limit.
This allows to take into account conservation of baryon
number, electric charge, and strangeness in the CE formulation
and, additionally,  energy conservation in the MCE. In
Section III the relevant formulas for the scaled
variance of multiplicity fluctuations are presented for different
statistical ensembles within the hadron-resonance gas model.  The
scaled variance of negative, positive and all charged hadrons
is then calculated along the chemical freeze-out line in the
temperature--baryon chemical potential plane. The fluctuations
of hadron multiplicities in central Pb+Pb (Au+Au) collisions are
presented for different collision energies from SIS  to
LHC.  The results for the GCE, CE, and MCE are compared.
In Section IV the statistical model predictions for
the scaled variances and multiplicity distributions
of negatively and positively charged hadrons
are compared with the preliminary NA49 data
of central Pb+Pb collisions in the SPS energy range.  A
summary, presented in Section V, closes the paper.
New features
of resonance decays within the MCE are discussed in Appendix A,
and the acceptance procedure for all charged hadrons is considered in Appendix B.

\section{Multiplicity Fluctuations in Statistical Models}
The mean multiplicities of positively, negatively and all charged
particles are defined as:
 \eq{
 \langle N_-\rangle \;=\; \sum_{i,q_i<0} \langle N_i\rangle\;,~~~~
 \langle N_{+}\rangle \;=\; \sum_{i,q_i>0} \langle
 N_i\rangle\;,~~~~
 \langle N_{ch}\rangle \;=\; \sum_{i,q_i\neq 0} \langle
 N_i\rangle\;,
 \label{pminch}
 }
where the average final state (after
resonance decays) multiplicities $\langle N_i\rangle$ are equal
to:
 \eq{\label{<N>}
 \langle N_i\rangle
 \;=\;
 \langle N_i^*\rangle + \sum_R \langle N_R\rangle \langle
 n_{i}\rangle_R\;.
 }
In Eq.~(\ref{<N>}), $N_i^*$ denotes the number of stable primary
hadrons of species $i$, the summation $\sum_R$ runs over all types
of resonances $R$,
and $\langle n_i\rangle_R \equiv \sum_r
b_r^R n_{i,r}^R$~ is the average  over resonance decay channels.
The parameters $b^R_r$ are the branching ratios of the $r$-th
branches, $n_{i,r}^R$ is the number of particles of species $i$
produced in resonance $R$ decays via a decay mode $r$. The index $r$
runs over all decay channels of a resonance $R$, with the
requirement $\sum_{r} b_r^R=1$.
In the GCE formulation of the hadron-resonance gas model
the mean number of stable
primary particles, $\langle N_i^* \rangle$, and the mean number of
resonances, $\langle N_R \rangle$, can be calculated as:
 \eq{\label{Ni-gce}
 \langle N_j\rangle \;\equiv\; \sum_{\bf p} \langle n_{{\bf p},j}\rangle
  \;=\; \frac{g_j V}{2\pi^{2}}\int_{0}^{\infty}p^{2}dp\; \langle
 n_{{\bf p},j}\rangle\;,
}
where $V$ is the system volume and $g_j$ is the degeneracy factor
of particle of the species $j$  (number of spin states). In the
thermodynamic limit, $V\rightarrow \infty$, the sum over the
momentum states can be substituted by a momentum integral.  The
$\langle n_{{\bf p},j} \rangle$ denotes the mean occupation number
of a single quantum state labelled by the  momentum vector ${\bf
p}$~,
 \eq{
 \langle n_{{\bf p},j} \rangle
 ~& = ~\frac {1} {\exp \left[\left( \epsilon_{{\bf p}j} - \mu_j \right)/ T\right]
 ~-~ \alpha_j}~, \label{np-aver}
  }
where $T$ is the system temperature, $m_j$ is the mass
of a particle $j$, $\epsilon_{{\bf p}j}=\sqrt{{\bf p}^{2}+m_j^{2}}$
is a single  particle energy. A value of $\alpha_j$ depends on quantum
statistics, it is $+1$ for bosons and $-1$ for fermions, while
$\alpha_j=0$ gives the Boltzmann approximation. The chemical
potential $\mu_j$ of a species $j$ equals to:
\eq{ \mu_j~=~q_j~\mu_Q~+~b_j~\mu_B~+~s_j~\mu_S ~,\label{chempot}}
where $q_j,~b_j,~s_j$ are the particle electric charge, baryon number, and
strangeness, respectively, while
$\mu_Q,~\mu_B,~\mu_S$ are the corresponding chemical potentials
which regulate the average values of these global conserved
charges in the GCE.
Eqs.~(\ref{Ni-gce}-\ref{chempot}) are valid in the GCE. In
the limit $V\rightarrow\infty$~, Eq.~(\ref{Ni-gce}-\ref{chempot})
are also valid for the CE and MCE, if
the energy density and conserved charge densities are the same in
all three ensembles. This is usually referred to as the
thermodynamical equivalence of all statistical ensembles. However, the
thermodynamical equivalence does not apply to fluctuations.

In statistical models a natural measure of multiplicity
fluctuations is the scaled variance of the multiplicity
distribution.
For negatively, positively, and all charged particles the
scaled variances read:
 \eq{
 \omega^- ~=~ \frac{\langle \left( \Delta N_- \right)^2
\rangle}{\langle N_-
  \rangle}~,~~~~
 \omega^+~ =~ \frac{\langle \left( \Delta N_+ \right)^2
\rangle}{\langle N_+
  \rangle}~,~~~~
 \omega^{ch}~ =~ \frac{\langle \left( \Delta N_{ch} \right)^2
\rangle}{\langle N_{ch}
  \rangle}~.
\label{omega-all}
 }
The variances in Eq.~(\ref{omega-all}) can be presented as a sum
of the correlators:
\eq{
\langle \left( \Delta N_- \right)^2 \rangle
 ~& =~
\sum_{i,j;~q_i<0,q_j<0} \langle \Delta N_i \Delta N_j
\rangle~,~~~~ \langle \left( \Delta N_+ \right)^2 \rangle
 ~ =~
\sum_{i,j;~q_i>0,q_j>0} \langle \Delta N_i \Delta N_j \rangle~,\nonumber \\
\langle \left( \Delta N_{ch} \right)^2 \rangle
 ~ &=~
\sum_{i,j;~q_i\neq 0,q_j\neq 0} \langle \Delta N_i \Delta N_j
\rangle~, \label{DNpm}
}
where $\Delta N_i\equiv N_i -\langle N_i\rangle$. The correlators
in Eq.~(\ref{DNpm}) include both the correlations between
primordial hadrons and those of final state hadrons due to the
resonance decays (resonance decays obey charge as well as energy-momentum conservation).

In the GCE the final state correlators can be calculated as \cite{Koch}:
 \eq{\label{corr-GCE}
  \langle \Delta N_i\,\Delta N_j\rangle_{g.c.e.}
  ~=~
  \langle\Delta N_i^* \Delta N_j^*\rangle_{g.c.e.}
  \;+\; \sum_R \left[ \langle\Delta N_R^2\rangle\;
  \langle n_{i}\rangle_R\;\langle n_{j}\rangle_R
  \;+\; \langle N_R\rangle\; \langle \Delta n_{i}\Delta n_{j}\rangle_R
  \right]~,
  }
where  $\langle \Delta n_i~\Delta n_j\rangle_R\equiv \sum_r b_r^R
n_{i,r}^R n_{j,r}^R~-~\langle n_i\rangle_R\langle n_j\rangle_R$~.
The  occupation numbers, $n_{{\bf p},j}$, of single quantum states (with
fixed projection of particle spin) are equal to
$n_{{\bf p},j}=0,1,\ldots,\infty$ for bosons  and
$n_{{\bf p},j}=0,1$ for fermions. Their average values are
given by Eq.~(\ref{np-aver}), and their fluctuations
read:
\eq{
 \langle~\left(\Delta n_{{\bf p},j}\right)^2~\rangle
~ \equiv ~ \langle \left( n_{{\bf p},j}~-~\langle
n_{{\bf p},j}\rangle\right)^2\rangle ~=~ \langle n_{{\bf p},j}\rangle \left(1~
+ ~\alpha_j ~\langle n_{{\bf p},j} \rangle\right)~\equiv~v^{ 2}_{{\bf p},j}~.
\label{np-fluc}
}
It is convenient to introduce a microscopic correlator,
$\langle \Delta n_{{\bf p},j}  \Delta n_{{\bf k},i} \rangle$,
which in the GCE has a simple form:
\eq{ \label{mcc-gce}
\langle \Delta n_{{\bf p},j}~  \Delta n_{{\bf k},i} \rangle_{g.c.e.}~=~
\upsilon_{{\bf p},j}^2\,\delta_{ij}\,\delta_{{\bf p}{\bf k}}~.
}
Hence there are no correlations between different particle species,
$i\neq j$, and/or between different momentum states, ${\bf p} \neq {\bf k}$.
Only the  Bose enhancement, $v_{{\bf p},j}^2>\langle n_{{\bf p},j}\rangle$ for
$\alpha_j=1$, and the Fermi suppression, $v_{{\bf p},j}^2<\langle
n_{{\bf p},j}\rangle$ for $\alpha_j=-1$, exist for fluctuations
of primary particles in the GCE. The correlator in Eq.~(\ref{corr-GCE}) can
be presented in terms of microscopic correlators (\ref{mcc-gce}):
\eq{ \label{dNidNj}
\langle \Delta N_j^* ~\Delta N_i^*~\rangle_{g.c.e.} ~=~
\sum_{{\bf p},{\bf k}}~\langle \Delta n_{{\bf p},j}~\Delta
n_{{\bf k},i}\rangle_{g.c.e.}~=~\delta_{ij}~\sum_{\bf p}~ v_{{\bf p},j}^2~.
 }
In the case of $i=j$ the above equation gives the  scaled variance
of primordial particles (before resonance decays) in the GCE.

In the MCE, the energy and conserved charges are fixed exactly
for each microscopic state of the system. This leads to two
modifications  in a comparison with the GCE. First, the additional terms appear
for the primordial microscopic
correlators in the MCE. They  reflect the (anti)correlations between
different particles, $i\neq j$, and different momentum levels,
${\bf p}\neq {\bf k}$, due to charge and energy
conservation in the MCE,
 \eq{\label{corr}
 &
 \langle \Delta n_{{\bf p},j}  \Delta n_{{\bf k},i} \rangle_{m.c.e.}
 ~=\;  \upsilon_{{\bf p},j}^2\,\delta_{ij}\,\delta_{{\bf p}{\bf k}}
 \;-\;  \frac{\upsilon_{{\bf p},j}^2v_{{\bf k},i}^2}{|A|}\;
 [\;q_iq_j M_{qq} + b_ib_j M_{bb} + s_is_j M_{ss} \nonumber
 \\
 &+ ~\left(q_is_j + q_js_i\right) M_{qs}~
 - ~\left(q_ib_j + q_jb_i\right) M_{qb}~
 - ~\left(b_is_j + b_js_i\right) M_{bs}\nonumber
 \\
 &+~ \epsilon_{{\bf p}j}\epsilon_{{\bf k}i} M_{\epsilon\epsilon}~-~
 \left(q_i \epsilon_{{\bf p}j} + q_j\epsilon_{{\bf k}i} \right)
 M_{q\epsilon}~
  +~ \left(b_i \epsilon_{{\bf p}j} + b_j\epsilon_{{\bf k}i} \right)
  M_{b\epsilon}~
  - ~\left(s_i \epsilon_{{\bf p}j} + s_j\epsilon_{{\bf k}i} \right) M_{s\epsilon}
 \;]\;,
 }
where $|A|$ is the determinant and $M_{ij}$ are the minors of the
following matrix,
 \eq{\label{matrix}
 A =
 \begin{pmatrix}
 \Delta (q^2) & \Delta (bq) & \Delta (sq) & \Delta (\epsilon q)\\
 \Delta (q b) & \Delta (b^2) & \Delta (sb) & \Delta (\epsilon b)\\
 \Delta (q s) & \Delta (b s) & \Delta (s^2) & \Delta (\epsilon s)\\
 \Delta (q \epsilon) & \Delta (b \epsilon) & \Delta (s \epsilon) & \Delta (\epsilon^2)
 \end{pmatrix}\;,
 }
with the elements, $\;\Delta (q^2)\equiv\sum_{{\bf p},j}
q_{j}^2\upsilon_{{\bf p},j}^2\;$, $\;\Delta (qb)\equiv \sum_{{\bf p},j}
q_{j}b_{j}\upsilon_{{\bf p},j}^2\;$, $\;\Delta (q\epsilon)\equiv \sum_{{\bf p},j}
q_{j}\epsilon_{{\bf p}j}\upsilon_{{\bf p},j}^2\;$, etc. The sum, $\sum_{{\bf p},j}$~,
means integration over momentum ${\bf p}$, and the summation over all
hadron-resonance species~$j$ contained in the model. The first term in the r.h.s. of Eq.~(\ref{corr})
corresponds to the microscopic correlator (\ref{mcc-gce}) in the
GCE. Note that a presence of the terms
containing a single particle energy,
$\epsilon_{{\bf p}j}=\sqrt{{\bf p}^{2}+m_j^{2}}$, in
Eq.~(\ref{corr}) is a consequence of  energy conservation. In
the CE, only charges are conserved,
thus the terms containing $\epsilon_{{\bf p}j}$
in Eq.~(\ref{corr}) are absent. The  $A$ in
Eq.~(\ref{matrix}) becomes then the $3\times 3$ matrix (see
Ref.~\cite{res}).  An important
property of the microscopic correlator method is that
the particle number fluctuations and the correlations in the MCE or CE,
although being different from those in the GCE, are expressed by
quantities calculated within the GCE. The microscopic
correlator (\ref{corr}) can be used  to calculate the primordial
particle correlator  in the MCE (or in the CE):
\eq{
  \langle \Delta N_{i} ~\Delta N_{j}~\rangle_{m.c.e.}
 &~= \sum_{{\bf p},{\bf k}}~\langle \Delta n_{{\bf p},i}~\Delta
 n_{{\bf k},j}\rangle_{m.c.e.}\;. \label{mc-corr-mce}
}

A second  feature of the MCE (or CE) is the modification
of the resonance decay contribution to the fluctuations
in comparison to the GCE (\ref{corr-GCE}).
In the MCE it reads:
 \eq{
 \langle \Delta N_i\,\Delta N_j\rangle_{m.c.e.}
 ~&=\; \langle\Delta N_i^* \Delta N_j^*\rangle_{m.c.e.}
  \;+\; \sum_R \langle N_R\rangle\; \langle \Delta n_{i}\; \Delta  n_{j}\rangle_R
 \;+\; \sum_R \langle\Delta N_i^*\; \Delta N_R\rangle_{m.c.e.}\; \langle n_{j}\rangle_R
  \; \nonumber
 \\
 &+\; \sum_R  \langle\Delta N_j^*\;\Delta N_R\rangle_{m.c.e.}\; \langle n_{i}\rangle_R
  \;+\; \sum_{R, R'} \langle\Delta N_R\;\Delta N_{R'}\rangle_{m.c.e.}
  \; \langle n_{i}\rangle_R\;
       \langle n_{j}\rangle_{R^{'}}\;.\label{corr-MCE}
 }
Additional terms in Eq.~(\ref{corr-MCE}) compared to
Eq.~(\ref{corr-GCE}) are due to the correlations
(for primordial particles) induced by
energy and charge conservations in the MCE.
The Eq.~(\ref{corr-MCE}) has the same
form in the CE \cite{res} and MCE, the difference between these two ensembles
appears because of different microscopic correlators (\ref{corr}).
The microscopic correlators of the MCE together with
Eq.~(\ref{mc-corr-mce}) should be used to calculate
  the correlators $\langle\Delta N_i^*
\Delta N_j^*\rangle_{m.c.e.}$~,~$\langle\Delta N_i^*\; \Delta
N_R\rangle_{m.c.e.}~$, $~\langle\Delta N_j^*\;\Delta
N_R\rangle_{m.c.e.}~$, $~\langle\Delta N_R\;\Delta
N_{R'}\rangle_{m.c.e.}$ entering in Eq.~(\ref{corr-MCE}) .

The microscopic correlators and the scaled variance are connected with the width of the multiplicity distribution.
It can be shown \cite{CLT} that in statistical models
the form of
the multiplicity distribution derived within any ensemble
(e.g. GCE, CE and MCE) approaches the Gauss distribution:
\begin{equation}\label{Gauss}
P_G(N) = \frac{1}{\sqrt{2 \pi ~\omega~\langle N \rangle}} ~\exp \left[ -
\frac{\left(N~-~\langle N \rangle \right)^2}{2 ~\omega ~\langle N \rangle} \right]~
\end{equation}
in the large volume limit i.e. $\langle N \rangle \rightarrow \infty$.
The width of this Gaussian, $\sigma = \sqrt{\omega~ \langle N \rangle}$, is determined by
the choice of the statistical ensemble,  while from the
thermodynamic equivalence of the statistical ensembles
it follows that the expectation value $\langle N \rangle$ remains the same.

\section{multiplicity fluctuations at chemical freeze-out }\label{sec-HG}
In this section we present the results of the hadron-resonance gas
for the scaled variances in the GCE, CE and MCE along the chemical
freeze-out line in central Pb+Pb (Au+Au) collisions for the whole
energy range from SIS to LHC.
Mean hadron multiplicities in heavy ion collisions at high
energies can be approximately fitted by the GCE
hadron-resonance gas model. The fit parameters are
the volume $V$, temperature $T$, chemical
potential $\mu_B$, and the strangeness
saturation parameter $\gamma_S$.
The latter  allows for non-equilibrium
strange hadron yields. A recent discussion of system size
and energy dependence of freeze-out parameters and comparison of
freeze-out criteria can be found in Refs.~\cite{FOP,FOC}. There are
several programs designed for the analysis of particle
multiplicities in relativistic heavy-ion collisions within the
hadron-resonance gas model, see e.g., SHARE \cite{Share},
THERMUS \cite{Thermus}, and THERMINATOR \cite{Therminator}.
The set of model parameters, $V,T,\mu_B$, and $\gamma_S$,
corresponds to the chemical freeze-out
conditions in heavy-ion collisions.
The numerical values and evolution of the model parameters
with the collision energy are taken from previous analysis of
multiplicities data.
The dependence of $\mu_B$ on the collision energy
is parameterized as \cite{FOC}:
$\mu_B \left( \sqrt{s_{NN}} \right) =1.308~\mbox{GeV}\cdot(1+
0.273~ \sqrt{s_{NN}})^{-1}~,$
where the c.m. nucleon-nucleon collision energy, $\sqrt{s_{NN}}$,
is taken in GeV units. The system is assumed
to be net strangeness free, $S=0$, and to have the
charge to baryon ratio of the initial colliding nuclei, $Q/B = 0.4$.
These two conditions define the system strange, $\mu_S$, and electric, $\mu _Q$,
chemical potentials.
For the chemical freeze-out condition we chose the average energy per particle,
$\langle E \rangle/\langle N \rangle = 1~$GeV \cite{Cl-Red}.
Finally, the strangeness saturation factor, $\gamma_S$,
is  parametrized \cite{FOP},
$ \gamma_S~ =~ 1 - 0.396~ \exp \left( - ~1.23~ T/\mu_B \right). $
This determines all parameters of the model.
In this paper an extended version of
THERMUS  framework \cite{Thermus} is used.
A numerical procedure is applied to meet the above
constraints simultaneously.
Other choices of the freeze-out parameters will be discussed in the next section.
The $T$, $\mu_B$, and $\gamma_S$ parameters used for different c.m. energies are given in Table~I.
Here, some further details
should be mentioned.
We use quantum statistics, but disregard the non-zero widths of resonances.
The thermodynamic limit
for the calculations of the scaled variance is assumed,
thus $\omega$ reaches its limiting value,
and volume $V$ is
not a parameter of our model calculations. We also do not consider explicitly momentum conservation
as
it can be shown that it completely drops out
in the thermodynamic limit \cite{CLT}.
Excluded volume corrections due to a hadron hard
core volume are not taken into account. They will be considered elsewhere \cite{ExclVol}.
The standard THERMUS particle
table includes all known particles and resonances up to a mass of about 2.5~GeV and
their respective  decay channels. Heavy resonances have not always well
established decay channels.
 We re-scaled the branching ratios given in THERMUS to unity,
where it was necessary
to ensure a global charge conservation.  Usually the resonance decays
are considered in a successive manner, hence, each resonance decays
into lighter ones until only stable particles are left. However, we need to implement
another procedure when different branches are
defined in a way that final states with only stable  hadrons are
counted.
This distinction does not affect mean quantities, but for
fluctuations it is crucial.
To make a correspondence with NA49 data, both strong
and electromagnetic decays should be taken into
account, while weak decays should be omitted.
\begin{table}[h!]
\begin{center}
\begin{tabular}{||c||c|c|c||c|c|c||c|c|c||c|c|c||}\hline
 $\sqrt{s_{NN}}$& $T$ &$\mu_B$&$\gamma_S$& \multicolumn{3}{c||}{ $\;\omega^{-}\;$} & \multicolumn{3}{c||}{
  $\;\omega^{+}\;$} & \multicolumn{3}{c||} { $\;\omega^{ch}\;$} \\ [0.5ex]
\hline [ GeV ] &[ MeV ]
&[ MeV ] & &\;GCE\; & \;CE\; & \;MCE\; & \;GCE\; & \;CE\; & \;MCE\; & \;GCE\; & \;CE\; & \;MCE\;  \\
[0.5ex] \hline\hline
$ 2.32 $ & 64.9  & 800.8 &0.641 &  1.025  &  0.777   &  0.578   &  1.020  &  0.116  &  0.086   &   1.048  &   0.403   &   0.300    \\
$ 4.86 $ & 118.5 & 562.2 &0.694 &  1.058  &  0.619   &  0.368   &  1.196  &  0.324  &  0.192   &   1.361  &   0.850   &   0.505    \\
$ 6.27 $ & 130.7 & 482.4 &0.716 &  1.069  &  0.640   &  0.346   &  1.203  &  0.390  &  0.211   &   1.431  &   0.969   &   0.524    \\
$ 7.62 $ & 138.3 & 424.6 &0.735 &  1.078  &  0.664   &  0.334   &  1.200  &  0.442  &  0.222   &   1.476  &   1.060   &   0.534    \\
$ 8.77 $ & 142.9 & 385.4 &0.749 &  1.084  &  0.683   &  0.328   &  1.197  &  0.479  &  0.230   &   1.504  &   1.126   &   0.541    \\
$ 12.3 $ & 151.5 & 300.1 &0.787 &  1.097  &  0.729   &  0.320   &  1.185  &  0.563  &  0.247   &   1.557  &   1.271   &   0.558    \\
$ 17.3 $ & 157.0 & 228.6 &0.830 &  1.108  &  0.768   &  0.318   &  1.174  &  0.637  &  0.263   &   1.593  &   1.393   &   0.576    \\
$ 62.4 $ & 163.1 & 72.7  &0.975 &  1.127  &  0.827   &  0.316   &  1.147  &  0.782  &  0.298   &   1.636  &   1.609   &   0.613    \\
$ 130  $ & 163.6 & 36.1  &0.998 &  1.131  &  0.827   &  0.313   &  1.141  &  0.805  &  0.305   &   1.639  &   1.631   &   0.618    \\
$ 200  $ & 163.7 & 23.4  &1.000 &  1.133  &  0.826   &  0.312   &  1.140  &  0.811  &  0.307   &   1.639  &   1.636   &   0.619    \\
$ 5500 $ & 163.8 & 0.9   &1.000 &  1.136  &  0.820   &  0.310   &  1.137  &  0.820  &  0.309   &   1.640  &   1.640   &   0.619    \\

\hline
\end{tabular}
\caption{The chemical freeze-out parameters $T$, $\mu_B$, $\gamma_S$, and
final state scaled variances in the GCE, CE, and MCE for central
Pb+Pb (Au+Au) collisions at different c.m. energies,
$\sqrt{s_{NN}}$.} \label{OmegaTableFinal}
\end{center}
\end{table}
Once a suitable set of thermodynamical parameters is determined
for each collision energy, the scaled variance of negatively, positively,
and all charged particles can be calculated using
Eqs.~(\ref{omega-all}-\ref{DNpm}). The
Eqs.~(\ref{corr-GCE}-\ref{dNidNj}) lead to the scaled variance in
the GCE, whereas Eqs.~(\ref{corr}-\ref{corr-MCE}) correspond to
the MCE (or CE) results. The $\omega^{-},~\omega^+,~\omega^{ch}$ in
different ensembles are presented in Table I for different
collision energies. The values of
$\sqrt{s_{NN}}$ quoted in Table~I correspond to the beam energies at
SIS (2$A$~GeV), AGS (11.6$A$~GeV), SPS ($20A$, $30A$, $40A$, $80A$, and
$158A$~GeV), colliding energies at RHIC ($\sqrt{s_{NN}}=62.4$~GeV,
$130$~GeV, and $200$~GeV), and LHC ($\sqrt{s_{NN}}=5500$~GeV).

\begin{figure}[ht!]
\begin{center}
 \epsfig{file=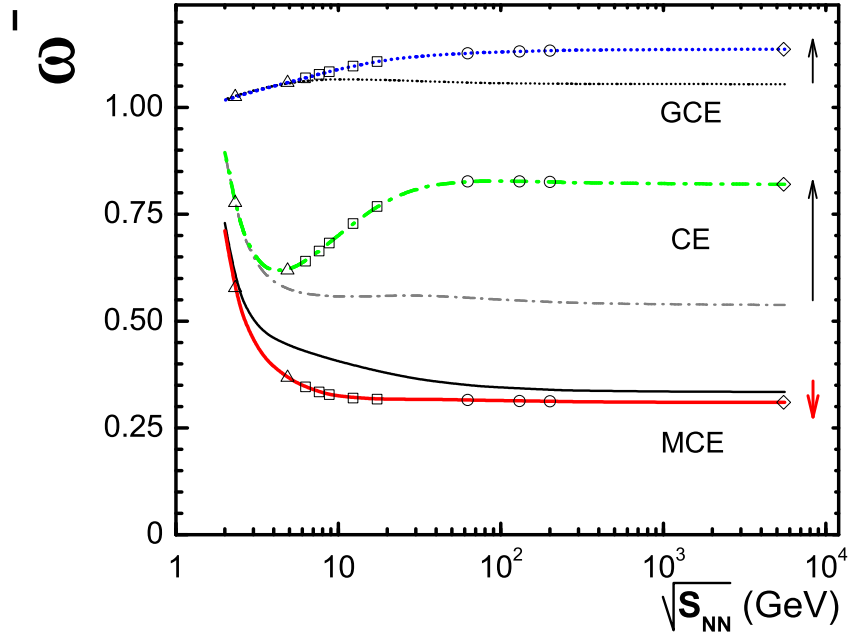,height=8.7cm}
 \caption{The scaled variances for negatively charged
particles, $\omega^-$, both primordial and final, along the
chemical freeze-out line for central Pb+Pb (Au+Au) collisions.
Different lines present the GCE, CE, and MCE results. Symbols at
the lines for final particles correspond to the
specific collision energies pointed out in Table I. The arrows
show the effect of resonance decays.} \label{omega_m}
\end{center}
\end{figure}

\begin{figure}[ht!]
\begin{center} \epsfig{file=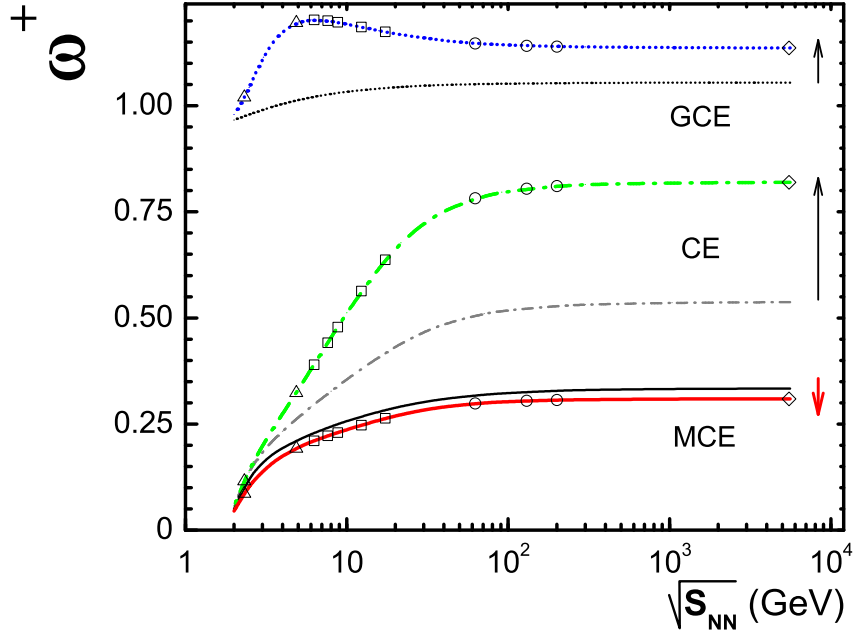,height=8.7cm}
 \caption{ The same as in Fig.~\ref{omega_m}, but for $\omega^+$.}
 \label{omega_p}
\end{center}
\end{figure}

\begin{figure}[ht!]
\begin{center} \epsfig{file=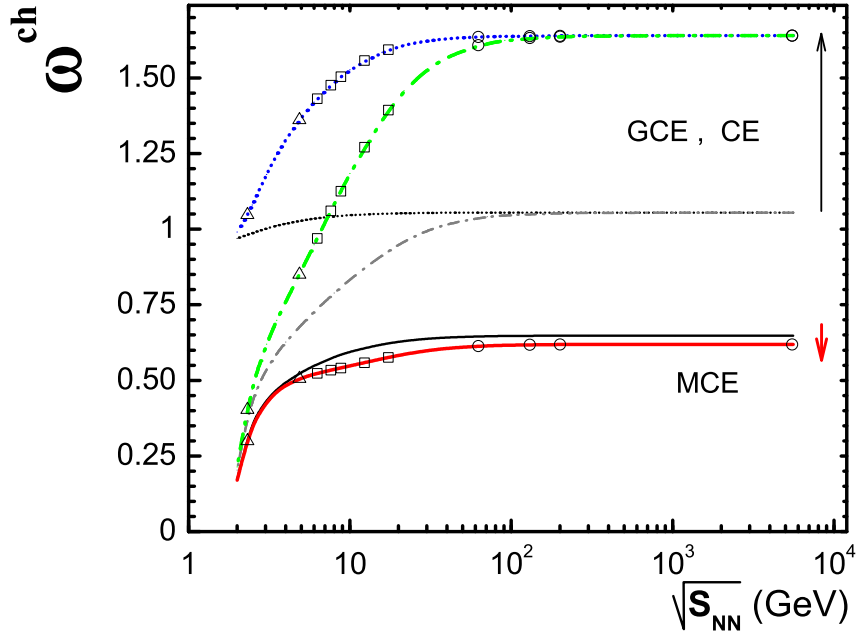,height=8.7cm}
 \caption{The same as in Figs.~\ref{omega_m} and \ref{omega_p}, but for $\omega^{ch}$.}
 \label{omega_ch}
\end{center}
\end{figure}

The mean multiplicities, $\langle N_i\rangle$, used for calculation
of the scaled variance
(see Eq.~(\ref{omega-all})) are given by Eqs.~(\ref{<N>}) and (\ref{Ni-gce})
and remain the same in all three ensembles. The
variances in Eq.~(\ref{omega-all}) are calculated using the
corresponding correlators $\langle \Delta N_i \Delta N_j \rangle$
in the GCE, CE, and MCE. For the calculations of
final state correlators the summation in Eq.~(\ref{corr-MCE}) should
include all resonances $R$ and $R^{\prime}$ which have particles of
the species $i$ and/or $j$ in their decay channels.
The resulting
scaled variances
are presented in Table I and shown in
Figs.~\ref{omega_m}-\ref{omega_ch}
as the functions of
$\sqrt{s_{NN}}$.

At the chemical
freeze-out of heavy-ion collisions, the Bose
effect for pions and resonance decays are important  and thus
(see also Ref.~\cite{res}):
$\omega^-_{g.c.e.}\cong 1.1$, $\omega^+_{g.c.e.}\cong 1.2$, and
$\omega^{ch}_{g.c.e.}\cong 1.4\div 1.6$, at the SPS energies.
Note that in the Boltzmann approximation and neglecting the resonance
decay effect one finds
$\omega^-_{g.c.e.}=\omega^+_{g.c.e.}=\omega^{ch}_{g.c.e.}=1$.

Some qualitative features of the results
should be mentioned. The effect of Bose and Fermi statistics is
seen in primordial values in the GCE. At low temperatures most of
charged hadrons are protons, and Fermi statistics dominates,
$\omega^{+}_{g.c.e.}, ~\omega^{ch}_{g.c.e.}<1$.
On the other hand, in the
limit of high temperature (low $\mu_B/T$) most charged hadrons are
pions and the effect of Bose statistics dominates, $\omega_{g.c.e.}^{\pm},~
\omega_{g.c.e.}^{ch}>1$. Along the chemical freeze-out line,
$\omega_{g.c.e.}^-$ is always slightly larger than 1, as $\pi^-$ mesons
dominate at both low and high
temperatures. The bump in $\omega^+_{g.c.e.}$ for final state particles
seen in Fig.~\ref{omega_p} at the small collision energies  is
due to a correlated production of proton and
$\pi^+$ meson from
$\Delta^{++}$
decays. This single resonance contribution
dominates in $\omega^+_{g.c.e.}$ at small collision energies
(small temperatures), but becomes relatively unimportant at the high
collision energies.

A minimum in $\omega_{c.e.}^{-}$ for final particles is seen in
Fig.~\ref{omega_m}. This is due to two effects. As the number of
negatively charged particles is relatively small, $\langle
N_-\rangle \ll \langle N_+\rangle$, at the low collision energies,
both the CE suppression and the resonance decay effect are small.
With increasing $\sqrt{s_{NN}}$, the CE effect alone leads to a
decrease of $\omega^-_{c.e}$, but the resonance decay effect only
leads to an increase of $\omega^-_{c.e}$. A combination of these
two effects, the CE suppression and the resonance enhancement,
leads to a minimum of $\omega^-_{c.e}$. As expected,
$\omega_{m.c.e.}<\omega_{c.e.}$, as an energy conservation further
suppresses the particle number fluctuations. A new feature of the
MCE  is the additional suppression of the fluctuations after
resonance decays. This is discussed in Appendix A.

\section{comparison with NA49 data}
\subsection{Centrality Selection}
The fluctuations in nucleus-nucleus collisions are studied on an
event-by-event basis: a given quantity is measured for each collision
and a distribution of this quantity is measured for a selected sample
of these collisions. It has been found that the fluctuations in the
number of nucleon participants give the dominant contribution to
hadron multiplicity fluctuations.
In the language of statistical
models, fluctuations of the number of nucleon participants
correspond to volume fluctuations caused by the variations
in the collision geometry.
Mean hadron multiplicities  are proportional (in the large volume limit)
to the volume, hence, volume fluctuations translate
directly to the multiplicity fluctuations.
Thus a comparison between  data and predictions of statistical
models should be performed for results which correspond to
collisions with a fixed number of nucleon participants.

Due to experimental limitations it is only possible to
approximately  measure
the number of participants of the projectile nucleus, $N_P^{proj}$,
in fixed target experiments (e.g. NA49 at the CERN SPS).
This is done in NA49 by measuring the energy deposited in a downstream Veto calorimeter.
A large fraction of this energy is due to projectile spectators $N_S^{proj}$.
Using baryon number conservation for the projectile nucleus
($A = N_P^{proj}+N_S^{proj}$) the number of projectile participants can be
estimated.
However, also a fraction of non-spectator particles, mostly protons and neutrons,
contribute to the Veto energy~\cite{NA49}.
Furthermore, the total number of nucleons participating in the
collision can fluctuate considerably even
for collisions with a fixed number of projectile participants
(see Ref.~\cite{Voka}).
This is due to fluctuations of the number of target participants.
The consequences of the asymmetry in an event selection depend on the dynamics of nucleus-nucleus
collisions (see Ref.~\cite{MGMG} for details).
Still, for the most central Pb+Pb collisions selected by the
number of projectile participants an increase of the scaled variance
can be estimated to be smaller than a few \% \cite{MGMG}
due to the target participant fluctuations.
In the following our predictions will be compared with the
preliminary NA49
data on the 1\% most central
Pb+Pb collisions at 20$A$-158$A$ GeV \cite{NA49}.
The number of projectile participants for these collisions is estimated to be larger than 193.

\subsection{Modelling of Acceptance}
In the experimental study of nuclear collisions at high energies
only a fraction of all produced particles
is registered.
Thus, the multiplicity distribution of the measured particles is expected to
be different from the distribution of all produced particles.
Let us consider the production of $N$ particles with the probability
$P_{4\pi}(N)$ in the full momentum space.
If particle detection is uncorrelated,
this means that the detection of one particle has no influence
on the probability to detect another one, the binomial distribution can be used.
For a fixed
number of produced particles $N$
the multiplicity distribution of accepted particles
reads:
\begin{align}\label{bin}
P_{acc}(n,N)~=~q^n(1-q)^{N-n}~\cdot \frac{N!}{n!(N-n)!}~,
\end{align}
where  $n\le N$ and  $q$ is the probability of a single particle to be accepted (i.e.
it is the ratio between mean multiplicity
of accepted and all hadrons).
Consequently one gets,
$\overline{n}=q~N~,~~
\overline{n^2}-\overline{n}^2=q(1-q)N~$,
where
$\overline{n^k}\equiv\sum_{n=0}^{N} n^k P_{acc}(n,N)~$, for $k=1,2,\ldots$~.
The probability distribution $P(n)$  of the accepted particles reads:
\eq{\label{W-acc}
P(n)~=~\sum_{N=n}^{\infty}P_{4\pi}(N)~P_{acc}(n,N)~.
}
The first two
moments of the distribution $P(n)$ are
calculated as:
\begin{align}\label{ac1}
\langle n \rangle & ~\equiv~\sum
_{N=0}^{\infty}P_{4\pi}(N)~ \sum_{n=0}^{N}
n~P_{acc}(n,N)~=~q \cdot \langle N\rangle~,\\
\langle n^2 \rangle & ~\equiv~\sum
_{N=0}^{\infty}P_{4\pi}(N)~ \sum_{n=0}^{N}
n^2~P_{acc}(n,N)~=~q^2\cdot\langle N^2\rangle~
+~q (1-q) \cdot \langle N \rangle ~,\label{ac2}
\end{align}
where ($k=1, 2,\ldots$)
\begin{align}\label{ac3}
\langle N^k \rangle ~ \equiv~ \sum_{N=0}^{\infty}
N^k~P_{4\pi}(N)~.
\end{align}
Finally, the scaled variance for the accepted particles can be
obtained:
\begin{align}\label{ac4}
\omega~\equiv~\frac{\langle n^2 \rangle~-~\langle n \rangle ^2}{\langle n\rangle}~  =~1~-~q~ +q\cdot \omega_{4\pi}~,
\end{align}
where $\omega_{4\pi}$ is the scaled variance of the $P_{4\pi}(N)$
distribution. The limiting behavior of $\omega$ agrees with the expectations.
In the large acceptance limit ($q \approx 1$) the distribution of
measured particles approaches the distribution in the full
acceptance.
For a very small acceptance ($q \approx 0$) the measured distribution
approaches the Poisson one independent of the shape of the
distribution in the full acceptance.

Model results on multiplicity fluctuations presented in Sec. III
correspond to an ideal situation when all final hadrons are
accepted by a detector. For a comparison with experimental data a
limited detector acceptance should be taken into account.  Even if
primordial particles at chemical freeze-out are only weakly
correlated in momentum space this would no longer be valid for
final state particles as resonance decays lead to momentum
correlations for final hadrons. In general, in statistical models,
the  correlations in momentum space are caused by resonance
decays, quantum statistics and the energy-momentum conservation
law, which is implied in the MCE. In this paper we neglect these
correlations and use Eqs.~(\ref{W-acc}) and (\ref{ac4}). This may
be approximately valid for $\omega^+$ and $\omega^-$, as most
decay channels only contain one positively (or negatively) charged
particle, but is certainly much worse for $\omega^{ch}$, for
instance due to  decays of neutral resonances into two charged
particles. In order to limit correlations caused by resonance
decays, we focus on the results for negatively and positively
charged hadrons. A discussion of the effect of resonance decays to
the acceptance procedure and a comparison with the data for
$\omega^{ch}$ are presented in Appendix B. An improved modelling
of the effect regarding the limited experimental acceptance will
be a subject of a future study.

 \subsection{Comparison with the NA49 Data for $\omega^-$ and $\omega^+$}
\begin{figure}[h!]
\epsfig{file=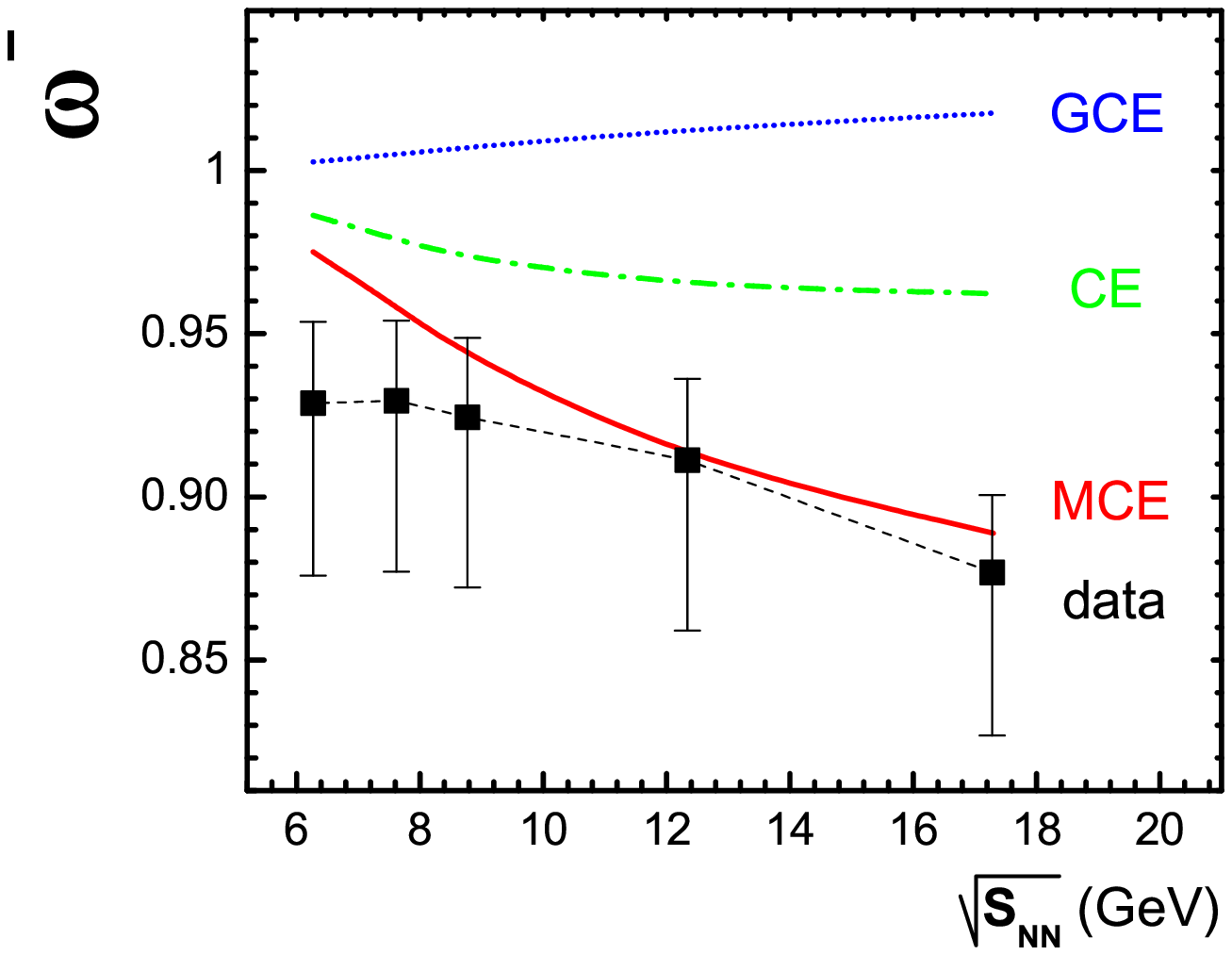,width=8.4cm}
\epsfig{file=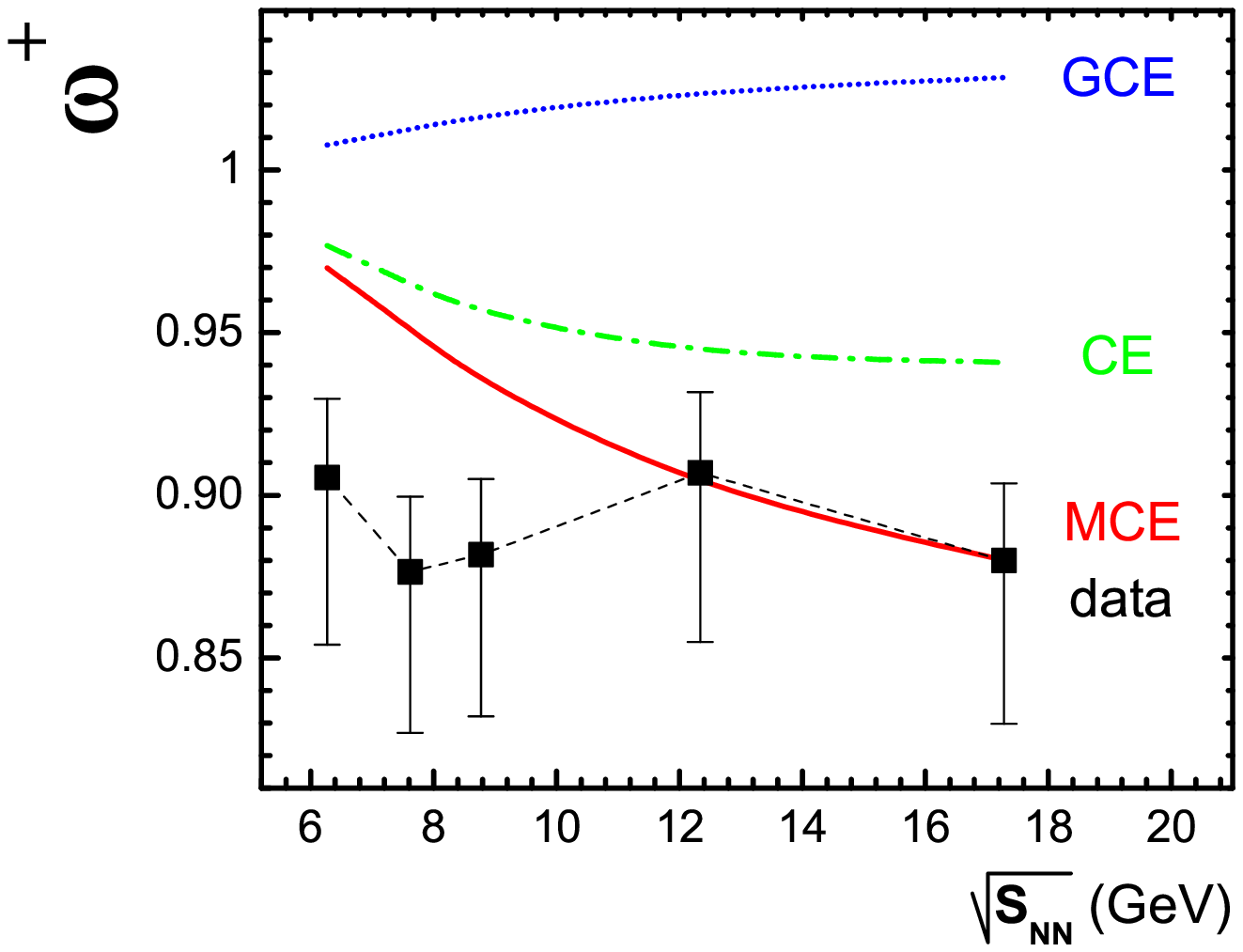,width=8.4cm}
 \caption{The scaled variances for negative (left)
and positive (right)
hadrons along the chemical freeze-out line for central Pb+Pb
collisions at the SPS energies. The points show the preliminary data
of NA49 \protect\cite{NA49}.
Total (statistical+systematic) errors are indicated. The statistical model
parameters $T$, $\mu_B$, and $\gamma_S$ at different SPS collision
energies are presented in Table~I.
Lines show the GCE, CE, and MCE results calculated
with the NA49 experimental acceptance according to
Eq.~(\ref{ac4}).}\label{stat-acc}
\end{figure}

Fig.~\ref{stat-acc} presents the
scaled variances $\omega^-$ and $\omega^+$ calculated
with Eq.~(\ref{ac4}).
The hadron-resonance gas calculations in the GCE, CE, and MCE
shown in Figs.~1 and 2 are used for the $\omega_{4\pi}^{\pm}$.
The NA49 acceptance used for the fluctuation measurements
is located in the forward hemisphere ($1<y(\pi)<y_{beam}$, where
$y(\pi)$ is the hadron rapidity calculated assuming pion mass
and shifted to the collision c.m. system~\cite{NA49}).
The
acceptance probabilities for positively and negatively charged
hadrons are approximately equal,
$q^+\approx q^-$, and the numerical values at
different SPS energies are:
$q^{\pm}=0.038,~0.063,~0.085,~ 0.131,~0.163$ at
$\sqrt{s_{NN}}=6.27,~7.62,~8.77,~12.3,~17.3$~GeV, respectively.
Eq.~(\ref{ac4}) has the following property: if $\omega_{4\pi}$
is smaller or larger than 1, the same inequality remains to be
valid for $\omega$ at any value of $0<q\le 1$. Thus
one has a strong qualitative difference between the predictions
of the statistical model valid for any freeze-out conditions and
experimental acceptances. The CE and MCE correspond to
$\omega_{m.c.e.}^{\pm}<\omega^{\pm}_{c.e.}<1$, and the GCE to
$\omega_{g.c.e.}^{\pm}> 1$.

From Fig.~\ref{stat-acc} it follows that the NA49 data for
$\omega^{\pm}$ extracted from 1\% of
the most central  Pb+Pb collisions at all
SPS energies are best described by the results of the hadron-resonance gas
model calculated within the MCE.
The data reveal even stronger suppression of the particle number
fluctuations.

\subsection{Dependence on the Freeze-out Parameters}
The relation
$E/N=1$~GeV \cite{Cl-Red} was used in our calculations to define the freeze-out conditions.
It does not give the best fit of the multiplicity data at each specific energy.
In this subsection we check the dependence of the statistical model
results for the scaled variances on the choice of the freeze-out parameters.
For this purpose we compare the results obtained for the
parameters used in this paper (model A) with two other sets
of the freeze-out parameters at SPS energies:
model B \cite{FOP} and model C \cite{pbm}.
The  corresponding values of $T$ and $\mu_B$ are presented in Fig.~5.

\begin{figure}[h!]
\label{T_muB}
\includegraphics[scale=0.7]{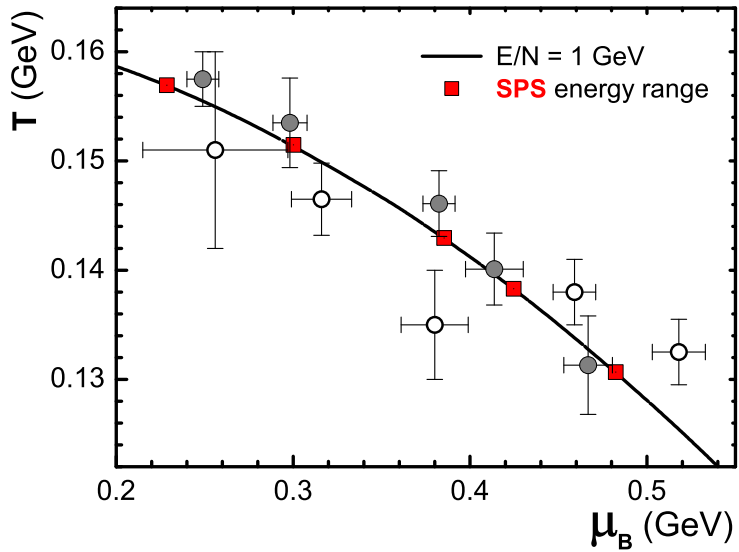}
\caption{Chemical freeze-out points in the $T$-$\mu_B$ plane for central
Pb+Pb collisions. The solid lines shows $\langle E \rangle /\langle N \rangle =
1$~GeV, the squares are from our parametrization (model A) and denote SPS beam
energies from $20A$~GeV (right) to $158A$~GeV (left), the full and open circles
are the best fit parameters from reference \cite{FOP} (model B)  and \cite{pbm} (model C),
respectively. }
\end{figure}

\begin{table}[h!]
\begin{center}\label{4piModelTable}
\begin{tabular}{||c||c|c|c||c|c|c||}\hline
$\sqrt{s_{NN}}$ &  \multicolumn{3}{c||}{ $\;\omega_{m.c.e.}^{-}\;$} & \multicolumn{3}{c||}{$\;\omega_{m.c.e.}^{+}\;$}  \\
[0.5ex] \hline
 [GeV]  & A & B & C &   A & B & C   \\
[0.5ex] \hline\hline
$ 6.27 $ & 0.346 & 0.345 & 0.361 & 0.211 & 0.214 & 0.210 \\
$ 7.62 $ & 0.334 & 0.334 & 0.347 & 0.222 & 0.225 & 0.221 \\
$ 8.77 $ & 0.328 & 0.330 & 0.330 & 0.230 & 0.232 & 0.236 \\
$ 12.3 $ & 0.320 & 0.318 & 0.325 & 0.247 & 0.249 & 0.248 \\
$ 17.3 $ & 0.318 & 0.317 & 0.321 & 0.263 & 0.264 & 0.259 \\
\hline
\end{tabular}
\caption{Final state scaled variances calculated in the MCE for 4 $\pi$
  acceptance using freeze-out conditions  A, B, and C.}
\end{center}
\end{table}

The scaled variances
$\omega_{m.c.e.}^-$ and $\omega_{m.c.e.}^+$ calculated in the full
phase space within the MCE vary by less than 1\% when changing the
parameter set.
In the NA49 acceptance the difference is almost completely
washed out. The differences are somewhat stronger in the GCE and CE, but
will not be considered here.

\subsection{Comparison of Distributions}

As discussed in Section II the multiplicity distribution in
statistical models in the full phase space and in the large volume
limit approaches a normal distribution. If the particle detection
is modelled by the simple procedure presented in Section IV B then
the results (\ref{ac1}-\ref{ac4}) are valid for any  form of the
full acceptance distribution $P_{4\pi} (N)$. In the following we
discuss  the properties of the multiplicity distribution in the
limited acceptance, $P(n)$, (\ref{W-acc}) and compare the
statistical model results in different ensembles with data on
negatively and positively charged hadrons.

For the Poisson distribution in the full acceptance
the summation in Eq.(\ref{W-acc})
leads also to the Poisson distribution in the acceptance
with the expectation value
$\langle n \rangle = q\langle N \rangle$:
\begin{equation}
P(n)~ =~ \sum_{N=n}^{\infty}\frac{\langle N\rangle^N \exp[-\langle N\rangle]}{N!}~\cdot~ \frac{q^n (1-q)^{N-n}~
N!}{n!(N-n)!}~=~
  \exp[- q ~\langle N \rangle] ~\frac{\left(q~
    \langle N \rangle \right)^n}{n!} ~.
\end{equation}
However,
the same does not hold true for summation in Eq.~(\ref{W-acc}) being applied
to other forms of the distribution $P(N)$.
In particular,
the normal distribution
(\ref{Gauss}) is transformed into the following:
\begin{equation}\label{model_acc}
P(n)~ = ~\sum_{N=n}^{\infty} P_{G}(N)~P_{acc}(n,N)~,
\end{equation}
which is not anymore the Gauss one.
It is enough to mention that a Gaussian is
symmetric around its mean value, while the distribution (\ref{model_acc}) is not.

\begin{figure}[h!]
\includegraphics[scale=0.65]{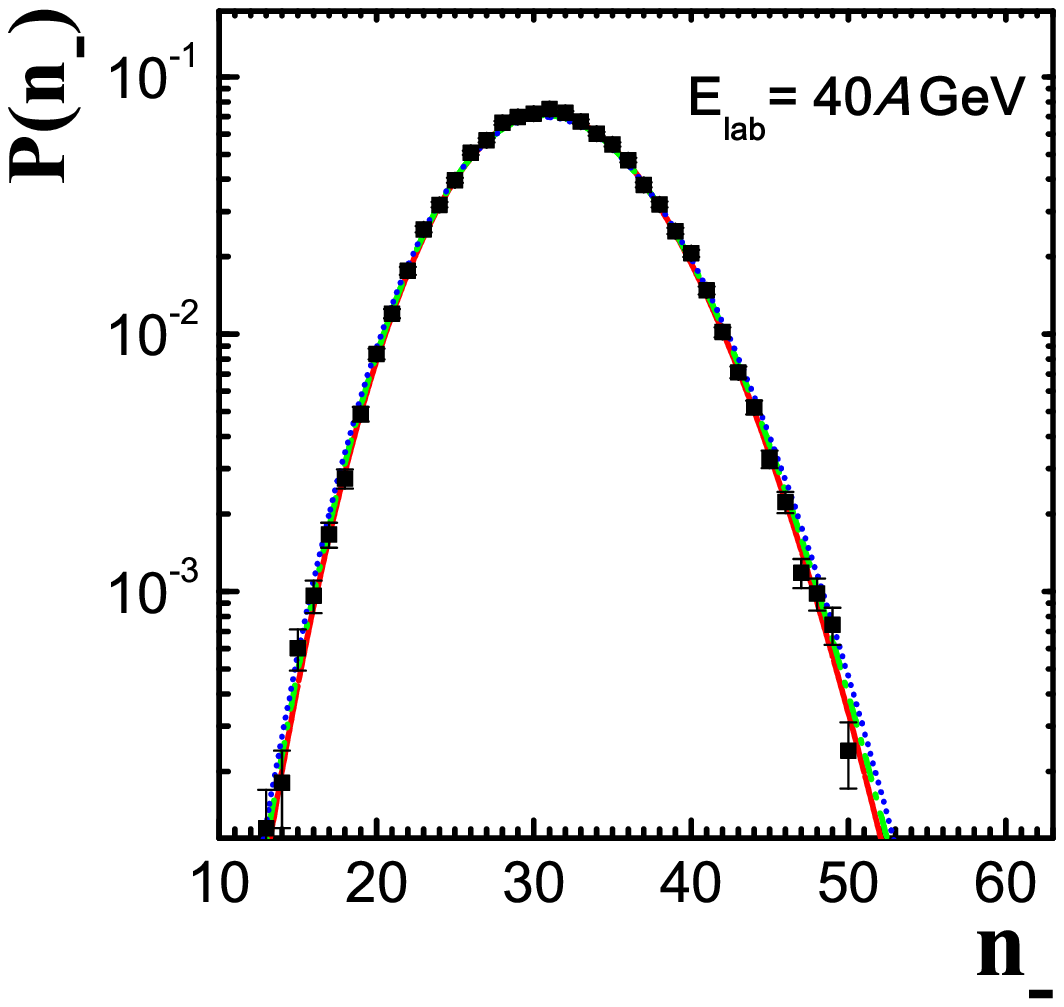}
\includegraphics[scale=0.65]{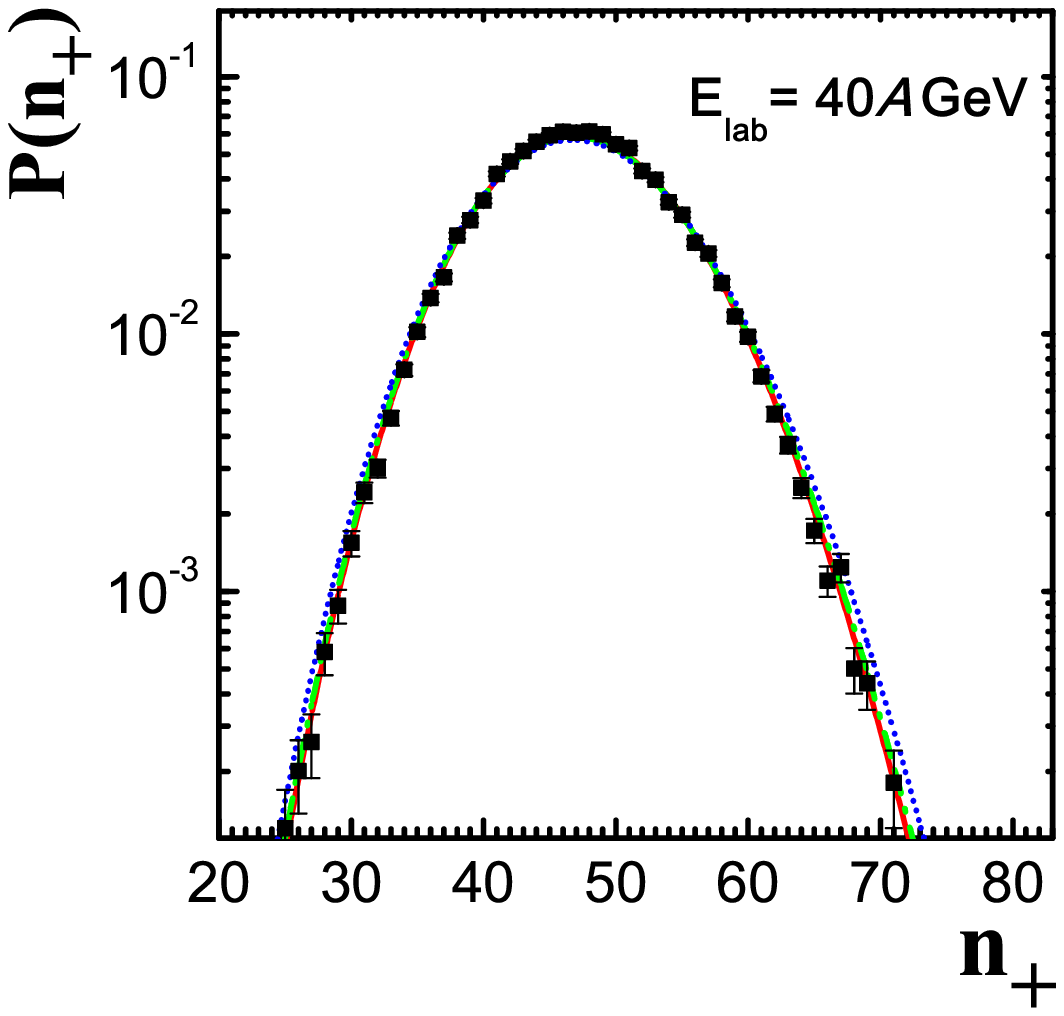}
\caption{
The  multiplicity distributions for negatively (left) and positively
(right) charged hadrons produced in central (1\%) Pb+Pb collisions
at 40$A$~GeV
in the NA49 acceptance \protect\cite{NA49}.
The preliminary experimental data (solid points)
of NA49 \protect\cite{NA49} are
compared with the prediction of the hadron-resonance gas
model obtained within different statistical ensembles,
the GCE (dotted lines), the CE (dashed-dotted lines) and
the MCE (solid lines).
}
\label{data1}
\end{figure}

The average number particles accepted by a detector is:
\begin{equation}
\langle n \rangle~ \equiv \sum_{n=0}^{\infty}n~P(n)~=~ q ~\langle N \rangle \equiv q ~\rho~ V~,
\end{equation}
where $\rho \equiv \langle N \rangle/V$ is the corresponding particle density.
 Hence, one can determine
the volume to be
\begin{equation}\label{FindVolume}
V ~= ~\frac{\langle N \rangle}{q ~ \rho}~.
\end{equation}
In the following
for each beam energy we adjust the volume to match the condition of
Eq.~(\ref{FindVolume}) for negatively ($V^-$) and positively ($V^+$) charged
yields, separately.
Note that values for the volume are about $10-20$\% larger than the ones
in \cite{FOP,pbm}, which were obtained using a much
less stringent centrality selection (here only the $1$\% most central data is
analyzed). We find that the $V^-$ and $V^+$ volume parameters deviate from each
other by less than 10\%.
Deviations of a similar magnitude are observed between the data
on hadron yield systematics and the hadron-resonance gas model fits.
Here we are only interested in the shape of multiplicity
distributions and do not attempt to optimize the model to fit simultaneously
yields of positively and negatively charged particles.
As typical examples
the multiplicity distributions for negatively and positively
charged hadrons produced in central Pb+Pb collisions
at 40$A$ GeV are shown in Fig.~\ref{data1}
at the SPS energy range.

\begin{figure}[h!]
\label{DistNegPlot}
\includegraphics[scale=0.8]{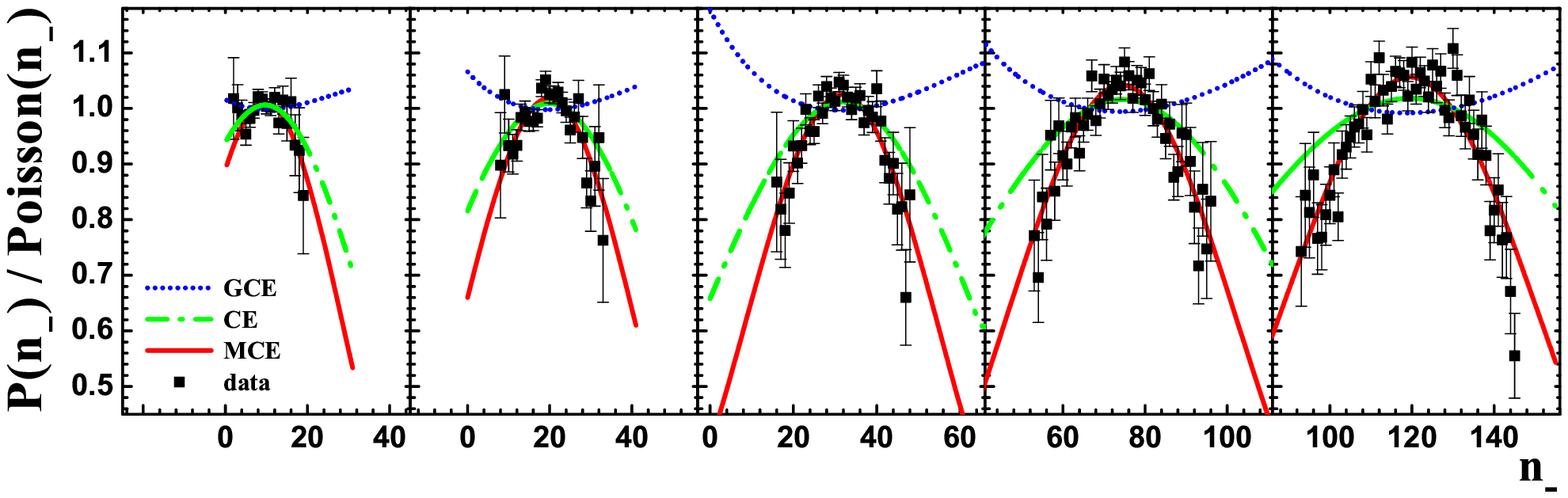}
\includegraphics[scale=0.8]{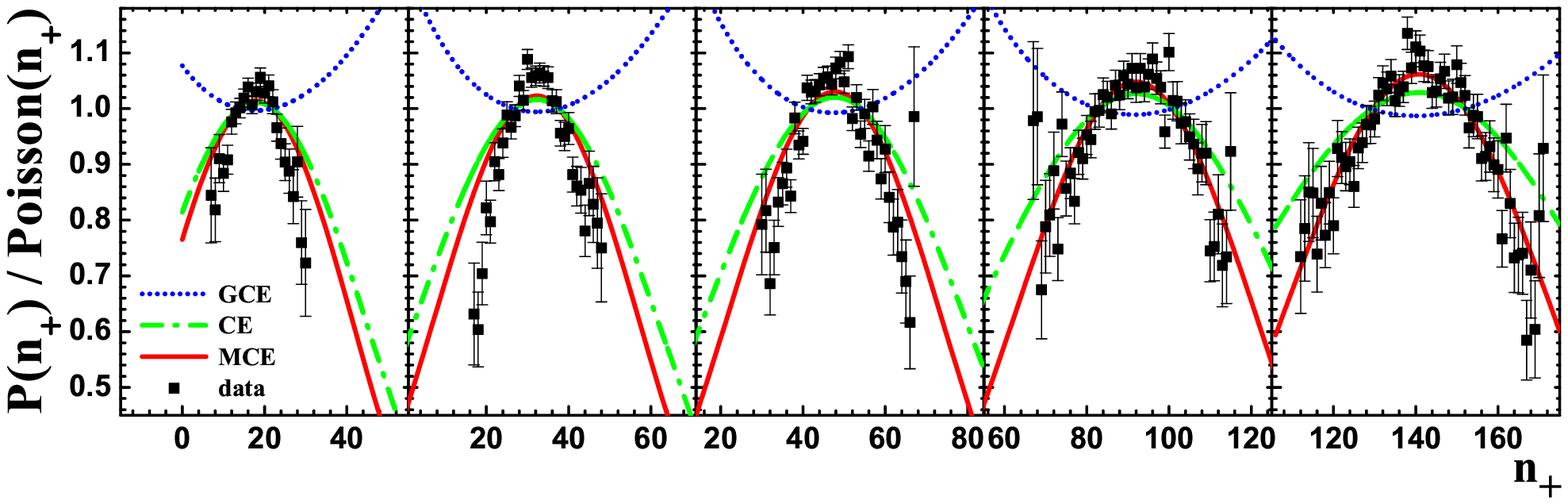}
\caption{
The ratio of the multiplicity distributions to Poisson ones for negatively
(upper panel) and positively (lower panel)
charged hadrons produced in central (1\%) Pb+Pb collisions
at 20$A$~GeV, 30$A$~GeV, 40$A$~GeV, 80$A$~GeV, and 158$A$~GeV
(from left to right)
in the NA49 acceptance \protect\cite{NA49}.
The preliminary experimental data (solid points)
of NA49 \protect\cite{NA49} are
compared with the prediction of the hadron-resonance gas
model obtained within different statistical ensembles,
the GCE (dotted lines), the CE (dashed-dotted lines), and
the MCE (solid lines).
}
\label{data2}
\end{figure}

The bell-like shape of the measured spectra is well
reproduced by the shape predicted by the statistical model.
In the semi-logarithmic plot differences between the data and
model lines obtained within different statistical ensembles
are hardly visible.
In order to allow for a detailed comparison of the distributions
the ratio of the data and the model distributions to the Poisson
one is presented in Fig.~\ref{data2}.

The results for negatively and positively charged hadrons
at 20$A$~GeV, 30$A$~GeV, 40$A$~GeV, 80$A$~GeV, and 158$A$~GeV
are shown separately.
The convex shape of the data reflects the fact that the
measured distribution is significantly narrower than the
Poisson one. This suppression of fluctuations
is observed for both charges, at all
five SPS energies and it is consistent with the results for the
scaled variance shown and discussed previously.
The GCE hadron-resonance gas results are broader than
the corresponding Poisson distribution. The ratio has a
concave shape.
An introduction of the quantum number conservation laws
(the CE results) leads to the convex shape and significantly
improves agreement with the data.
Further improvement of the agreement is obtained by the
additional introduction of the energy conservation law
(the MCE results). The measured spectra surprisingly well
agree with the MCE predictions.

\subsection{Discussion}

High resolution of the NA49 experimental data allows
to distinguish between multiplicity fluctuations
expected in hadron-resonance gas model for different
statistical ensembles. The measured spectra clearly
favor predictions of the micro-canonical ensemble.
Much worse description is obtained for the canonical
ensemble and
a strong disagreement is seen considering the grand canonical
one.
All calculations are performed in the thermodynamical
limit which is a proper approximation for the considered
reactions.
Thus these results should be treated as a first observation
of the recently predicted \cite{CE} suppression of multiplicity
fluctuations due to conservation laws
in relativistic gases in the large volume
limit.

A validity of the micro-canonical description is surprising even
within the framework of the statistical hadronization model used
in this paper. This is because in the calculations the parameters
of the model (e.g. energy, volume, temperature and chemical
potential) were assumed to be the same in all collisions. On the
other hand, significant event-by-event fluctuations of these
parameters may be expected. For instance, only a part of the total
energy is available for the hadronization process. This part
should be used in the hadron-resonance gas calculations while the
remaining energy is contained in the collective motion of matter.
The ratio between the hadronization and collective energies may
vary from collision to collision and consequently increase the
multiplicity fluctuations.

The agreement between the data and the MCE predictions is even
more surprising when the  processes which are beyond the
statistical hadronization model are considered. Examples of these
are jet and mini-jet production, heavy cluster formation, effects
related to the phase transition or instabilities of the
quark-gluon plasma. Naively all of them are expected to increase
multiplicity fluctuations and thus lead to a disagreement between
the data and the MCE predictions. A comparison of the data with
the models which include these processes is obviously needed for
significant conclusions. Here we consider only one example.

In Ref.~\cite{ood} a non-monotonic dependence of the relative
fluctuations,
\eq{\label{Re}
R_e~=~\frac{(\delta S)^2/S^2}{(\delta E)^2/E^2}~,
}
has been suggested as a signal for the onset of deconfinement.
Here $S$ and $E$ denote the system entropy and thermalized energy
at the early stage of collisions, respectively. This prediction
assumes event-by-event fluctuations of the thermalized energy,
which results in the fluctuations of the produced entropy. The
ratio of the entropy to energy fluctuations (\ref{Re}) depends on
the equation of state and thus on the form of created matter. The
$R_e$ is approximately independent of collision energy and equals
about $0.6$ in pure hadron or quark-gluon plasma phases. An
increase of the $R_e$ ratio up to its maximum value, $R_e\approx
0.8$, is expected \cite{ood} in the transition domain. Anomalies
in energy dependence of the hadron production properties measured
in central Pb+Pb collisions \cite{NA49-1} indicate \cite{mg2} that
the transition domain is located at the low CERN SPS energies,
from 30$A$ to 80$A$ GeV. Thus an anomaly in the energy dependence
of multiplicity fluctuations is expected in the same energy domain
\cite{ood}.

In any case the fluctuations of the thermalized energy will lead
to
additional multiplicity fluctuations (``dynamical fluctuations'').
The resulting contribution to the scaled variance can be
calculated to be:
\eq{\label{dyn}
\omega^{-}_{dyn}~=~R_e~\langle n_-\rangle~ \frac{(\delta
E)^2}{E^2}~.
}
The above assumes that the mean particle multiplicity is
proportional to the early stage entropy.
In order to perform a quantitative estimate of the effect the
fluctuations of the energy of produced particles were calculated
within the HSD \cite{hsd} and UrQMD \cite{urqmd} string-hadronic
models. For central (impact parameter zero) Pb+Pb collisions in
the energy range from 30A to 80A~GeV we have obtained,
$\sqrt{(\delta E)^2}/E \leq 0.03$. The number of
accepted negatively charged particles is $\langle
n_-\rangle\approx 30$ at $40A$ GeV (see Fig.~\ref{data2}). Thus,
an increase of the $\omega$ due to
the ``dynamical fluctuations'' estimated by Eq.~(\ref{dyn}) is
$\omega^{-}_{dyn} \leq 0.02$ for $R_e=0.6$, and it is smaller than the
experimental error of the preliminary NA49 data of about 0.05 \cite{NA49}.
In particular, an additional increase due to the phase transition,
$\Delta \omega^{-}_{dyn}\approx 0.005$, for $R_e=0.8$, can be hardly observed.

In conclusion, the predicted \cite{ood} increase of the
scaled variance of the multiplicity distribution due to
the onset of deconfinement  is too small to be observed
in the current data.  These data neither confirm nor
refute the interpretation \cite{mg2} of the measured \cite{NA49-1} anomalies
in the energy depedence of other hadron production properties
as due to the onset of deconfinement at the CERN SPS energies.

More differential data on multiplicity fluctuations and
correlations are required for further tests of the validity of the
statistical models and observation of possible signals of the
phase transitions.
The experimental resolution in a measurement of the enhanced
fluctuations due to the onset of deconfinement can be increased by
increasing acceptance. For example, $\omega_{dyn}^-\propto
\langle n_-\rangle\propto q$. The present aceptance of  NA49
at 40$A$ GeV is about $q\cong 0.06$ and it can be increased up to
about $q\cong 0.5$ in the future studies. This will give a chance
to observe, for example, the dynamical fluctuations discussed in
Ref. \cite{ood}. The observation of the MCE suppression effects of the
multiplicity 
fluctuations by NA49 was possible only because a selection of a
sample of collisions without projectile spectators. This selection
seems to be possible only in the fixed target experiments. In the
collider kinematics nuclear fragments which follow the beam
direction can not be measured.

On the model side a further study is needed to improve description
of the effect of
the limited experimental acceptance.
Further on, a finite volume of hadrons is expected
to lead to a reduction of
the particle number fluctuations \cite{ExclVol}. A quantitative
estimate of this effect is needed.

\section{Summary}
The hadron multiplicity fluctuations in relativistic
nucleus-nucleus collisions have been predicted in the statistical
hadron-resonance gas model within the grand canonical, canonical, and
micro-canonical ensembles
in the thermodynamical limit.
The microscopic correlator method has been extended to include three
conserved charges -- baryon number, electric charge, and
strangeness -- in the canonical ensemble, and additionally an  energy
conservation in the micro-canonical ensemble.
The analytical formulas are used for the resonance
decay contributions to the correlations and fluctuations.
The scaled variances of negatively, positively, and
all charged particles for primordial and final state hadrons have been
calculated at the chemical freeze-out in central Pb+Pb (Au+Au)
collisions for different collision energies from SIS  to
LHC.  A comparison
of the multiplicity distributions and the scaled variances
with the preliminary NA49 data on  Pb+Pb collisions
at the SPS energies has been done for the samples of about
1\% of most central
collisions selected by the number of projectile participants.
This selection allows to eliminate effect of fluctuations of
the number of nucleon participants.
The effect of the limited experimental acceptance was taken
into account by use of the uncorrelated
particle approximation.

The   measured multiplicity distributions are significantly
narrower than the Poisson one and
allow to
distinguish between model results derived within different
statistical ensembles. The data surprisingly well agree
with the expectations for the micro-canonical ensemble
and exclude the canonical and grand-canonical ensembles.
Thus this is a first experimental observation of the
predicted suppression of the multiplicity fluctuations
in relativistic gases in the thermodynamical limit due
to conservation laws.

\begin{acknowledgments}
We would like to thank  F. Becattini, E.L.~Bratkovskaya,
K.A.~Bugaev, A.I.~Bugrij, W.~Greiner, A.P.~Kostyuk,
I.N.~Mishustin, St.~Mr\'owczy\'nski, L.M.~Satarov, H.~St\"ocker,
and O.S.~Zozulya for numerous discussions. We thank  E.V.~Begun
for help in the preparation of the manuscript. The work was
supported in part by US Civilian Research and Development
Foundation (CRDF) Cooperative Grants Program, Project Agreement
UKP1-2613-KV-04, Ukraine-Hungary cooperative project ¹ M/101-2005,
and Virtual Institute on Strongly Interacting Matter (VI-146) of
Helmholtz Association, Germany.

\end{acknowledgments}

\appendix
\section{Resonance decays in the MCE}
\begin{figure}[ht!]
\begin{center}
\epsfig{file=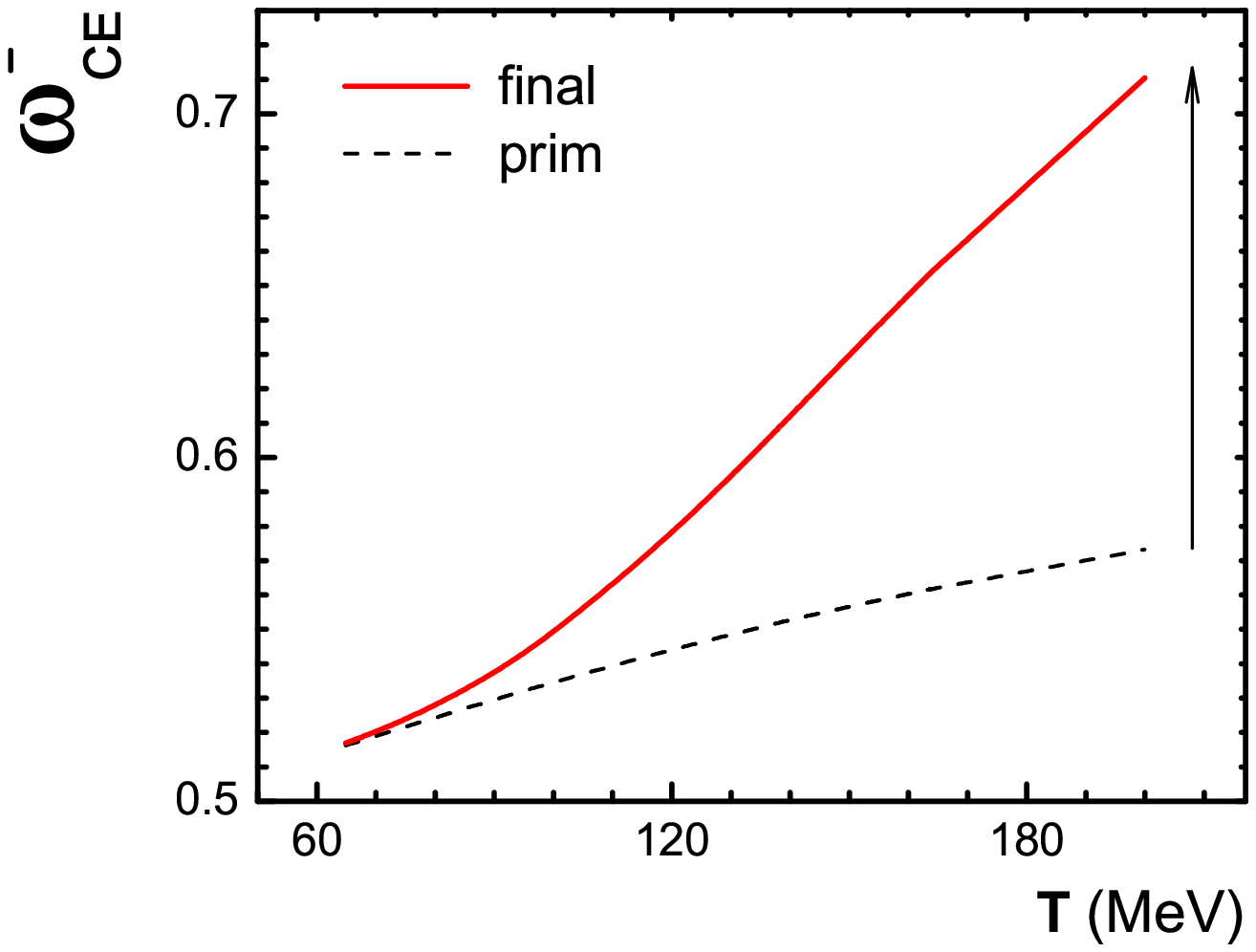,width=8.4cm}
\epsfig{file=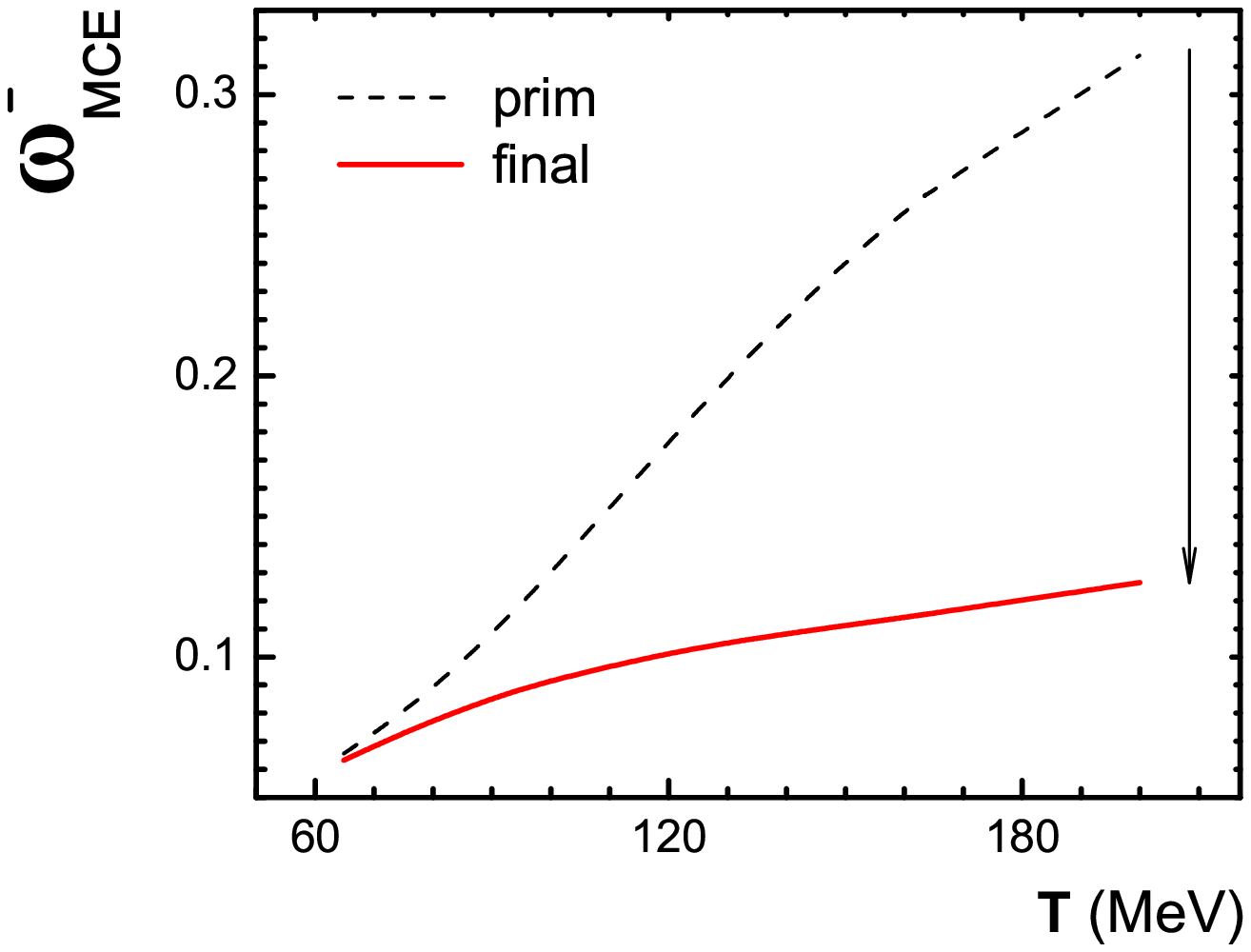,width=8.4cm}
 \caption{The scaled variance for negatively charged
particles, $\omega^-$, in the toy $(\pi^+,\pi^-,\rho^0)$-system
with $Q=0$ in the CE (left) and MCE (right) as a function of
temperature.  The temperature interval corresponds to that
presented in Table I. The dotted lines correspond to primordial
$\pi^-$-meson fluctuations, the solid lines -- to those after
$\rho^0$ decays. At small $T$ the $\rho^0$ contribution to the
pion number fluctuations is negligible, and it increases
with $T$. The contribution from $\rho^0$ decays to $\omega^-$ is
positive in the CE and negative in the MCE (see the text for
details).
} \label{pi-rho}

\end{center}
\end{figure}

A comparison of the primordial
scaled variances with those for final hadrons
demonstrates that the fluctuations generally increase
after resonance decays in the
GCE and CE (see more details in Ref.~\cite{res}), but they
decrease in the MCE.
In order to understand this effect let us
consider a toy model $(\pi^+,\pi^-,\rho^0)$-system with a zero net
charge, $Q=0$. Due to this last condition there is a full symmetry
between positively and negatively charged pions, and thus $\omega^+=\omega^-$.
Each $\rho^0$-meson decays into a $\pi^+\pi^-$-pair with 100\%
probability, i.e. $b_1^{\rho}=1$ and $\langle
n_-\rangle_{\rho^0}=1$. The predictions of the CE and MCE for
$(\pi^+,\pi^-,\rho^0)$-system are shown in Fig.~\ref{pi-rho}.
One observes that $\rho^0$ decays lead to an
enhancement of $\omega^-$ in the CE, and to its suppression in the
MCE.
In the CE one finds from Eqs.~(\ref{<N>}) and (\ref{corr-MCE}) for
$(\pi^+,\pi^-,\rho^0)$-system:
 \eq{
  \langle N_-\rangle
 ~ = ~ \langle
 N_{\pi^-}^* \rangle~+~
  \langle N_{\rho^0}\rangle\;,~~~~
 \langle \left(\Delta N_-\right)^2\rangle_{c.e.}
 ~ = ~ \langle\left(\Delta
 N_{\pi^-}^*\right)^2\rangle_{c.e.}~+~
 \langle\left(\Delta N_{\rho^0}\right)^2\rangle_{c.e.}\;.
 \label{pi-rho-CE}
 }
Note that the average multiplicities, $\langle N_{\pi^-}^* \rangle$
and $\langle N_{\rho^0}\rangle$, remain the same in the CE,
and the MCE.
From Eq.~(\ref{pi-rho-CE}) it follows:
\eq{
\omega_{c.e.}^-~=~\omega_{c.e.}^{-*}~\left[\frac{\langle
N_{\pi^-}^*\rangle
~+~(\omega_{c.e.}^{\rho^0}/\omega_{c.e.}^{-*})\cdot \langle
N_{\rho^0}\rangle}{\langle N_-\rangle}\right]~.
\label{omega-pi-rho-CE}
}
The $\omega_{c.e.}^{-*}$ is essentially smaller than 1 due to the
strong CE suppression (see Fig.~\ref{pi-rho}, left).
On the other hand, there is no CE
suppression for $\rho^0$ fluctuations,
$\omega_{c.e.}^{\rho^0}=\omega_{g.c.e.}^{\rho^0}\cong 1$.
Therefore, one
finds that $\omega_{c.e.}^{\rho^0}/\omega_{c.e.}^{-*}>1$, and from
Eq.~(\ref{omega-pi-rho-CE}) it immediately follows,
$\omega_{c.e.}^-~>~\omega_{c.e.}^{-*}$. Note that
$\omega_{g.c.e.}^{-*}\cong\omega_{g.c.e.}^{\rho^0}\cong 1$,
thus there is no enhancement of $\omega_{g.c.e.}^-$ due to $\rho^0$ decays.
In the MCE the multiplicity $\langle N_{-}\rangle$ remains the
same as in Eq.~(\ref{pi-rho-CE}).
The variance $\langle
\left(\Delta N_-\right)^2\rangle_{m.c.e.}$ is, however, modified
because of the anti-correlation between primordial $\pi^{-*}$ and
$\rho^0$ mesons in the MCE. From Eq.~(\ref{corr-MCE}) one finds
for our $(\pi^+,\pi^-,\rho^0)$-system,
\eq{
\langle \left(\Delta N_-\right)^2\rangle_{m.c.e.}
 ~ = ~ \langle\left(\Delta
 N_{\pi^-}^*\right)^2\rangle_{m.c.e.}~+~
  \langle\left(\Delta N_{\rho^0}\right)^2\rangle_{m.c.e.}~+~
  2~ \langle\Delta N_{\pi^-}^*~ \Delta N_{\rho^0}\rangle_{m.c.e.}~.
 \label{pi-rho-MCE}
 }
The last term in Eq.~(\ref{pi-rho-MCE}) appears due to
energy conservation in the MCE (it is absent in the CE).
This term is evidently negative,
which means that an anti-correlation occurs. A large (small) number of
primordial pions, $\Delta N_{\pi^-}^*>0~ (<0)$, requires a small
(large)  number of $\rho^0$-mesons, $\Delta N_{\rho^0} <0~(>0)$,
to keep the total energy fixed. Anti-correlation between
primordial pions and $\rho^0$-mesons makes the $\pi^-$ number fluctuations
smaller after resonance decays, i.e. $\omega_{m.c.e.}^- <
\omega_{m.c.e.}^{-*}$, as depicted in Fig.~\ref{pi-rho} (right). The
same mechanism works in the MCE for the full hadron-resonance gas.

\section{Acceptance effect for All Charged Particles}

In order to better understand an influence of the
momentum correlation due to resonance decays on
the multiplicity
fluctuations we define a toy model. Let us assume that there are two kinds of
particles produced. The first kind ($N$) is either stable or originates from
decay channels which contain only one particle of the type we are set to
investigate, while the second kind ($M$) produces 2 particles
of the selected type. This is described by the (unknown) probability
distribution $P_{4\pi} \left(N,M\right)$. We further assume that for both
types of particles, $N$ and $M$, separately the acceptance procedure defined by Eq.~(\ref{bin}) is
applicable. We also assume that once  particle $M$ is inside the experimental
acceptance, both its decay products will be so as well. Hence, the average number
of observed particles will be:
\begin{equation}
\langle n \rangle ~=~ \sum_{N=0}^{\infty} \sum_{M=0}^{\infty}~ P_{4\pi}
\left(N,M\right)~ \sum_{n=0}^{N} \sum_{m=0}^{M}~ (n+2~m) ~P_{acc} \left(n,N
\right)  P_{acc} \left(m,M \right)~.
\end{equation}
This leads immediately to:
\begin{equation}
\langle n \rangle ~=~ q \cdot \Big[  \langle N \rangle + 2~\langle M \rangle
\Big]~.
\end{equation}
One finds the second moment,
\begin{equation}
\langle n^2 \rangle ~=~ \sum_{N=0}^{\infty} \sum_{M=0}^{\infty} ~P_{4\pi}
\left(N,M\right)~ \sum_{n=0}^{N} \sum_{m=0}^{M}~ (n+2m)^2 ~P_{acc} \left(n,N
\right)  P_{acc} \left(m,M \right)~.
\end{equation}
Making use of the relation (\ref{ac2})
one obtains:
\begin{align}
\langle n^2 \rangle ~=~ q  \left( 1-q\right) \cdot \langle N \rangle ~+~ q^2
\cdot \langle N^2\rangle ~+~ 4 q^2 \cdot \langle N  M \rangle
~+~ 4  q  \left( 1-q\right)\cdot \langle M \rangle ~+~ 4q^2 \cdot \langle M^2
\rangle~.
\end{align}
Thus, for the scaled variance it follows:
\begin{equation}\label{ac6}
\omega ~\equiv~ \frac{\langle n^2 \rangle -\langle n \rangle^2}{\langle n
  \rangle} ~=~ q \cdot \omega_{4\pi} ~+~ \left(1-q \right) \cdot \Bigg[ \frac{\langle N \rangle
    + 4~\langle M \rangle}{\langle N \rangle + 2~\langle M \rangle} \Bigg]~,
\end{equation}
where $\omega_{4\pi}$ is obtained from the case $q=1$ and corresponds to the
original distribution $P_{4\pi} \left(N,M\right)$. For the second limiting
case of Eq.~(\ref{ac6}), $q\rightarrow 0$, one finds a scaled variance which
corresponds to that of two uncorrelated Poisson distributions with means
$q\langle N \rangle$ and $q\langle M \rangle$, respectively.
In this case, all primordial correlations due to energy and charge conservation or Bose (Fermi)
statistics are lost, but
particles produced by resonances of type $M$ are still detected in pairs.
\begin{figure}[h!]
\epsfig{file=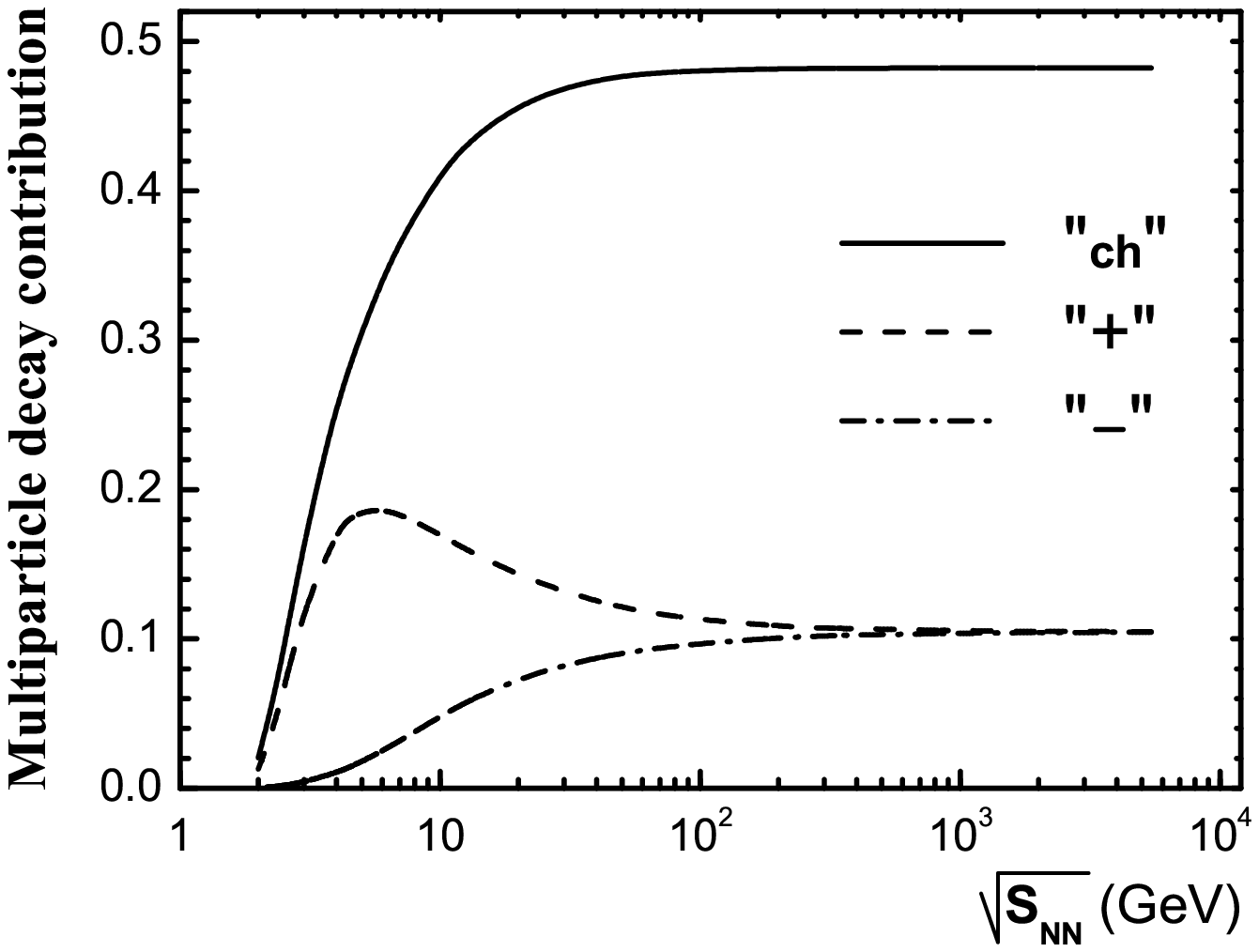,width=8.4cm}
\epsfig{file=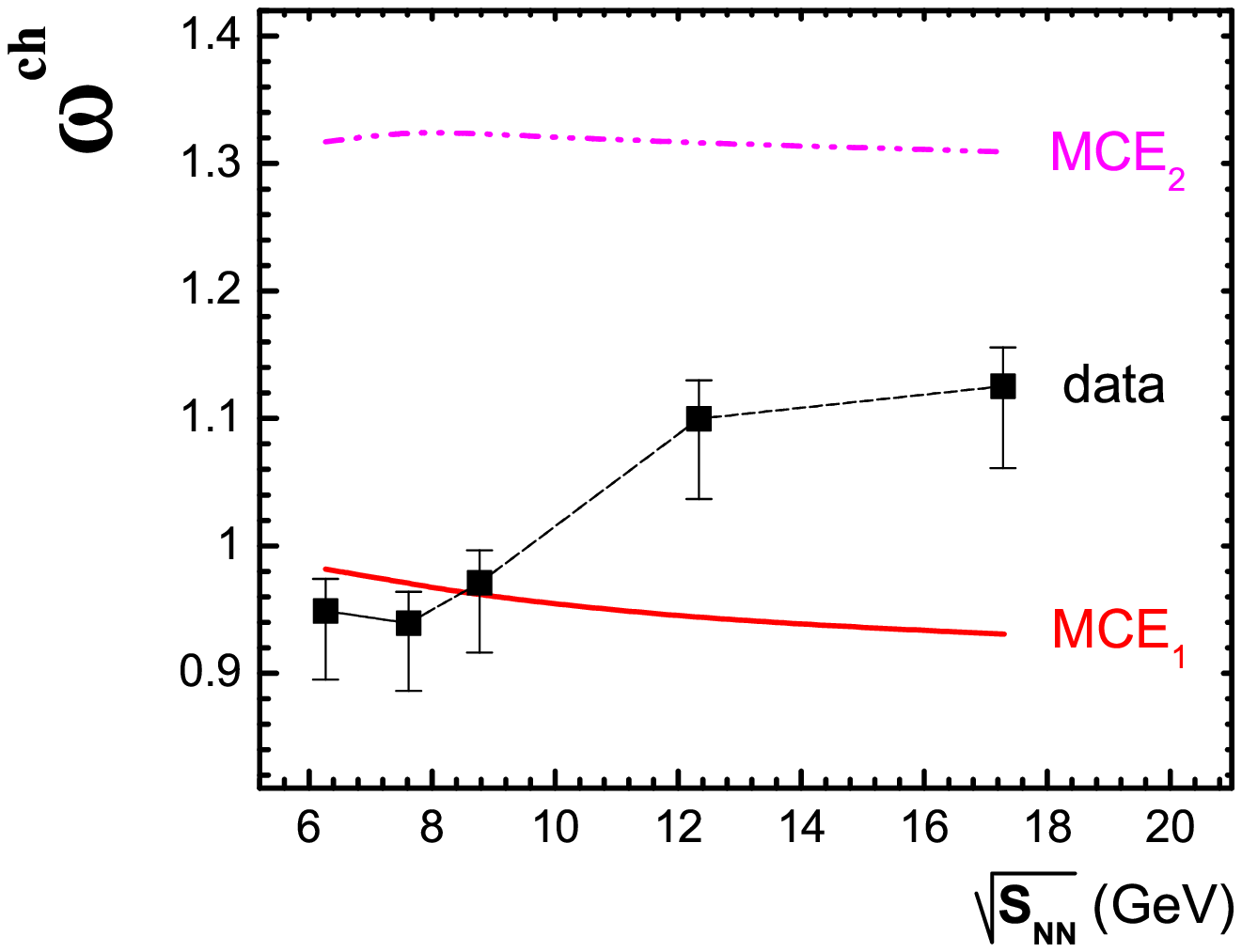,width=8.4cm}
\caption{{\it Left.} Fraction of total yield originating from resonance decays with 2 or
  more positively(`+'), negatively(`-'), or all  charged(`ch'),
  particles.  Parameters are taken from parametrization in section III.
 {\it Right.}  The scaled variances for all charged
hadrons along the chemical freeze-out line for central Pb+Pb
collisions at the SPS energies. The points show the preliminary data
of NA49. The statistical model
parameters $T$, $\mu_B$, and $\gamma_S$ at different SPS collision
energies are presented in Table I. The lines show the MCE
 results calculated with the NA49 experimental acceptance
according to Eq.~(\ref{ac4}) (lower line) and Eq.~(\ref{ac5}) (
upper line).}\label{DC}
\end{figure}
In general case, by $k$ we denote  the fraction of particles
originating from decays (always 2
relevant daughters) of particle kind $M$,  hence,
\begin{equation}
\langle N \rangle~ =~ \left( 1-k\right) ~\langle N_{tot} \rangle~, \qquad
\langle M \rangle~ = ~\frac{k}{2} ~ \langle N_{tot} \rangle~.
\end{equation}
Finally, one finds for the scaled variance:
\begin{equation}\label{ac5}
\omega  ~=~ q \cdot \omega_{4\pi} ~+~ \left(1-q \right) \left(1+k \right)~.
\end{equation}
From the hadron-resonance gas model we
can estimate the fraction $k$ of the final yield which originates from decays of
resonances into 2 (or more) charged particles.
From Fig.~\ref{DC} (left) we find the fraction of the charged
particle yield $k$ to be from 35\% ($20$~AGeV) to 45\% ($158$~AGeV) in the SPS energy
range (and about 10\% for positively and negatively charged
particles).
For the definition of decay channels see section III.
Examples of two-particle decay channels: $\rho^0 \rightarrow \pi^+ +
  \pi^-$, would  be counted as two particle decay in `ch', but neither in `+' nor in `-',
  $\Delta^{++} \rightarrow p + \pi^+$, would contribute to `ch' and `+', but not
  to `-`.
The assumption that both decay products are detected is certainly
not justifiable for small values of the total acceptance $q$, hence Eq.~(\ref{ac5})
overestimates the effect.
However, this consideration will give a useful upper
bound (see Fig.~\ref{DC}, right).
The typical width of decays is comparable to the
width of the acceptance window, therefore, about half of all decays will leave
one (or both) decay product missing. Yet the same 50\% will be contributed from decays
whose parents are outside the acceptance but contribute to the final
yield. Hence, one expects no change in average multiplicity, but a sizable effect on
fluctuations.

\end{document}